\documentclass [prx,floatfix,superscriptaddress,twocolumn]{revtex4-2}   	

\usepackage{graphicx}				
\usepackage{afterpage}	
\usepackage{amssymb}
\usepackage{amsmath}
\usepackage[version=3]{mhchem}
\usepackage{float}
\usepackage[flushleft]{threeparttable}
\usepackage{color}
\usepackage{endnotes}

\makeatletter
\newcommand{\colorcaption}[2][]{%
	\begingroup%
	\renewcommand{\@caption@fignum@sep}{ (color online). }%
	\caption[#1]{#2}%
	\endgroup%
}

\begin{document}			
	
\title{Violation of generalized fluctuation theorems in adaptively driven steady states: Applications to hair cell oscillations}
	
\author{Janaki Sheth}
\affiliation{Department of Physics and Astronomy, UCLA, Los Angeles California, 90095-1596, USA}	
\author{Dolores Bozovic}
\affiliation{Department of Physics and Astronomy, UCLA, Los Angeles California, 90095-1596, USA}
\affiliation{California NanoSystems Institute, UCLA, Los Angeles California, 90095-1596, USA}

\author{Alex J. Levine}
\affiliation{Department of Physics and Astronomy, UCLA, Los Angeles California, 90095-1596, USA}
\affiliation{Department of Chemistry and Biochemistry, UCLA, Los Angeles California, 90095-1596, USA}
\affiliation{Department of Computational Medicine, UCLA, Los Angeles California, 90095-1596, USA}	

\begin{abstract}
The spontaneously oscillating hair bundle of sensory cells in the inner ear is an example of a stochastic, nonlinear oscillator driven by internal active processes.  Moreover, this internal activity is adaptive -- its power input depends on the current state of the system.  We study fluctuation dissipation relations in such adaptively-driven, nonequilibrium limit-cycle oscillators.  We observe the expected violation of the well-known, equilibrium fluctuation-dissipation theorem (FDT), and verify the existence of a generalized fluctuation-dissipation theorem (GFDT) in the non-adaptively driven model of the hair cell oscillator. This generalized fluctuation theorem requires the system to be analyzed in the co-moving frame associated with the mean limit cycle of the stochastic oscillator.  We then demonstrate, via numerical simulations and analytic calculations, that the adaptively-driven dynamical hair cell model violates both the FDT and the GFDT. We go on to show, using stochastic, finite-state, dynamical models, that such a feedback-controlled drive in stochastic limit cycle oscillators generically violates both the FDT and GFDT.  We propose that one may in fact use the breakdown of the GFDT as a tool to more broadly look for and quantify the effect of adaptive, feedback mechanisms associated with driven (nonequilibrium) biological dynamics. 
\end{abstract}

\date{\today}	

\maketitle

\section{Introduction}
Biology is replete with nonequilibrium systems that expend energy to maintain cyclic steady-state dynamics. Examples include the chemical networks underlying circadian rhythms, activity patterns in neuronal networks, and cardiac rhythmogenesis~\cite{Biological-Clocks1960,Goldbeter1995,Mori1996,Goldbeter2002, Izhikevich2007,Schwab2010}. The inner ear provides another striking example of such dynamics, for it contains an internal active amplifier that allows the auditory system to detect nanoscale displacements ~\cite{Hudspeth2008, Reichenbach14, Martin1999}. In a quiet environment, the inner ear can moreover generate spontaneous otoacoustic emissions, which are metabolically sensitive, indicating the presence of an internal oscillatory instability that necessitates an energy source. Beyond its innate importance to the understanding of sensory neuroscience, the auditory system provides an experimentally tractable substrate in which to study nonequilibrium fluctuation theorems. In this work, we explore a generalization of the standard fluctuation dissipation theorem, and test it in theoretical models of inner ear dynamics.

An integral part of the vertebrate peripheral auditory system, hair cells of the inner ear~\cite{LeMasurier05} transduce mechanical displacements imposed by the incoming pressure waves into electrical signals. This process is mediated by direct mechanical gating, as specialized ion channels open in response to minute lateral deflections of the stereovilli ~\cite{Vollrath07} -- columnar structures protruding from the hair cells' apical surface and interconnected by tip links. When a hair bundle is deflected by an incoming sound, motion as small as a few {\AA}ngstroms leads to an increase in the tension exerted on the tip links and hence the opening of the transduction channels. Furthermore, hair cells demonstrate a number of adaptation processes that are key in maintaining this exquisite sensitivity. While the biophysical mechanisms behind their internal activity are not entirely known, a number of myosin motor species have been implicated in hair cells of different species, including Myosin 1C. These molecular motors climb along the internal actin filaments, and are believed to be connected to the tip links, thus providing a mechanism capable of continually adapting to incoming sounds and maintaining the optimal tension required for sensitive detection~\cite{Gillespie2004}.
\begin{figure}
	\includegraphics[width=1\linewidth]{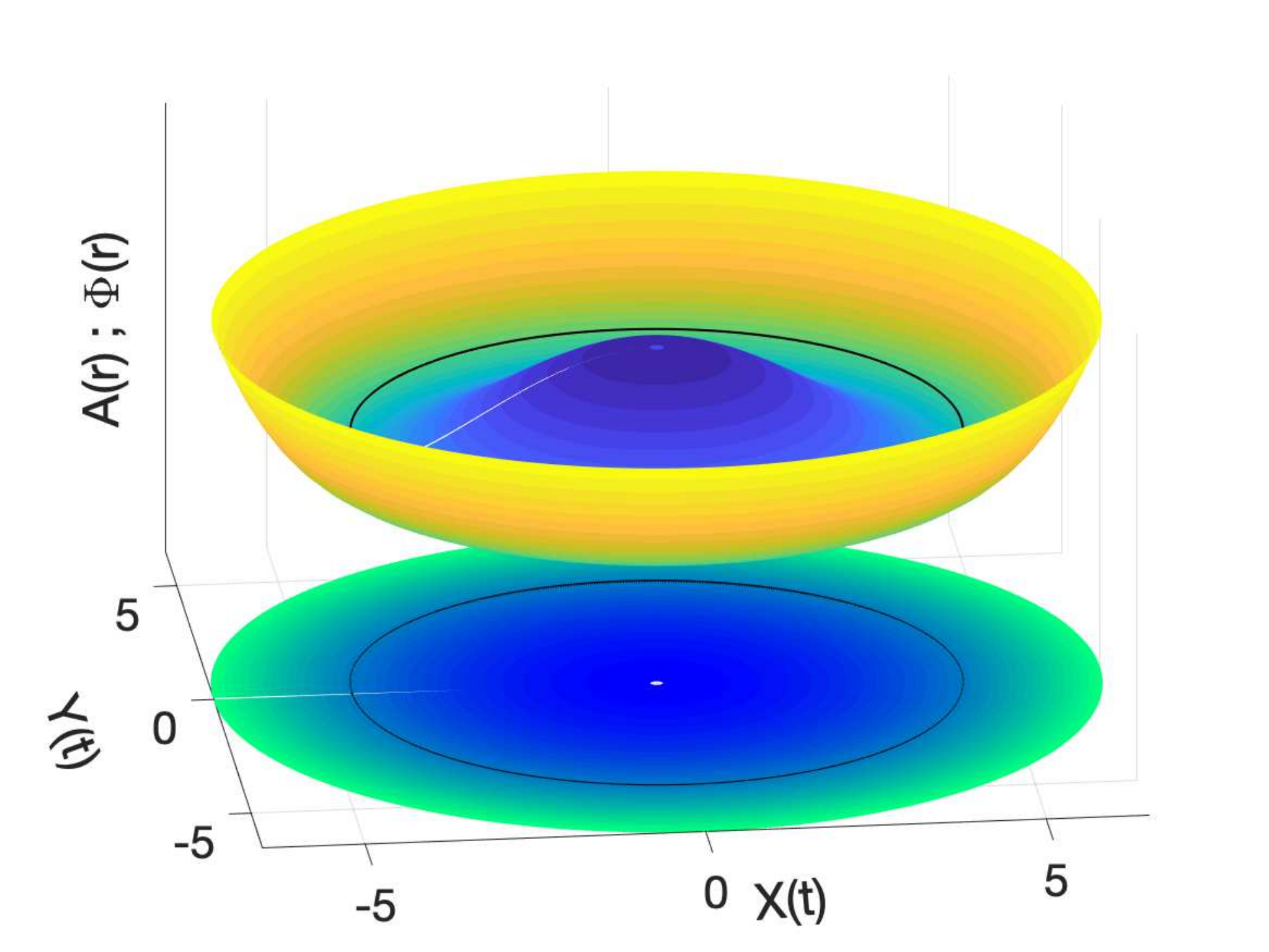}
	\colorcaption{The deterministic, unforced Hopf limit cycle (black curve) of radius $R_0$ sits in the azimuthally symmetric minimum potential region of $\Phi(r)$ as defined in Eq. \ref{scalar-hopf} and is driven by the curl of $A(r)$ given in Eq. \ref{vector-hopf}. The colormap for the three-dimensional $\Phi(r)$ runs from dark blue ($r = 0$) to yellow ($r = 7$). The magnitude of the vector potential $A(r)$ is shown as a colored disc which varies from dark blue ($r = 0$) to light green ($r = 7$). 
		\label{fig:Scalar_potential}}
\end{figure}

The mechanical feedback loop between the myosin motors and displacements of the stereovilli has important consequences for hair bundle dynamics. It 
allows for an unstable dynamical regime in which the bundle responds to mechanical input like a spring with a negative spring constant~\cite{Martin00}. In this regime, the bundle undergoes active oscillations even in the absence of incoming pressure waves due to the active feedback between motor activity and bundle displacement. 
Since that endogenous drive depends on the deflections of the bundle \textit{i.e.} on the state of our biological system, it provides a direct example of adaptive control of a nonequilibrium steady state. We note that similar examples of adaptation are found in a number of biological systems, including cellular regulation~\cite{Nguyen2012} and bacterial chemo-sensing~\cite{Lan2012}.

Biological systems are generally noisy, due to the thermal fluctuations of their constituent elements. Consistently with this, hair cells  encounter stochasticity from a number of sources. As the hair bundles are immersed in a fluid environment, surrounding Brownian motion leads to fluctuations of the stereovili on the order of a nanometer ~\cite{Nadrowski2004}. Force fluctuations resulting from stochastic myosin motor activity are also present and 
may contribute colored-noise~\cite{Nadrowski2004}. Finally, the membrane potential of the cell body, which is coupled to hair bundle motility,  
fluctuates due to ion channel clatter and shot noise in the ionic transport  through
transmembrane channels~\cite{Lauger1975}.  As a result, the limit cycle oscillations of the hair cell bundle are innately noisy and thus provide a window into the basic nonequilibrium statistical mechanics of a noisy limit cycle oscillator. Further, they operate under homeostatic control where the drive maintaining the nonequilibrium steady state responds to the internal state of the system.

Spontaneous oscillations of the hair bundle have been studied experimentally by direct measurements performed \textit{in vitro} on preparations of the amphibian sacculus ~\cite{Martin00, Martin2003, Hudspeth2008}.  Based on these experiments, a set of robust mathematical models of the active oscillations have been developed, comprising of nonlinear differential equations of varying degrees of complexity~\cite{Rami2014, Nadrowski2004, Martin2003}. The simplest model that captures the essential phenomena is a two-dimensional dynamical 
system that undergoes a supercritical Hopf bifurcation to the limit cycle (oscillatory) state~\cite{Hudsepth2014,Sheth2018}.   
In this manuscript, we use  the stochastic normal form equation for the Hopf bifurcation to model the spontaneously oscillating state of the hair bundle, in order to study fluctuation theorems associated with noisy nonequilibrium systems.  

It is well known that fundamental equilibrium fluctuation theorems can fail in nonequilibrium steady states.  In fact, the breakdown of the standard fluctuation 
dissipation theorem (FDT)~\cite{Callen1951} has been used as a way to characterize the nonequilibrium  steady state of cytoskeletal networks~\cite{Mizuno2007}.   More recently, there has been 
a new exploration of fluctuation theorems applicable to nonequilibrium steady states~\cite{Seifert2012}.  Here we show that, as expected, the driven hair bundle violates the standard, 
equilibrium FDT~\cite{Martin2001}, but does obey a generalized fluctuation dissipation theorem (GDFT), as predicted by the new body of work on fluctuation theorems in nonequilibrium steady states. This
agreement with the GDFT, however is predicated on the drive being nonadaptive, meaning that the power input from the drive does not depend on the current state of the system.  Once
we include this feature related to the adaptive control of the hair bundle oscillations, we obtain new violations of the nonequilibrium GFDT.
\begin{figure}
\includegraphics[width=0.8\linewidth]{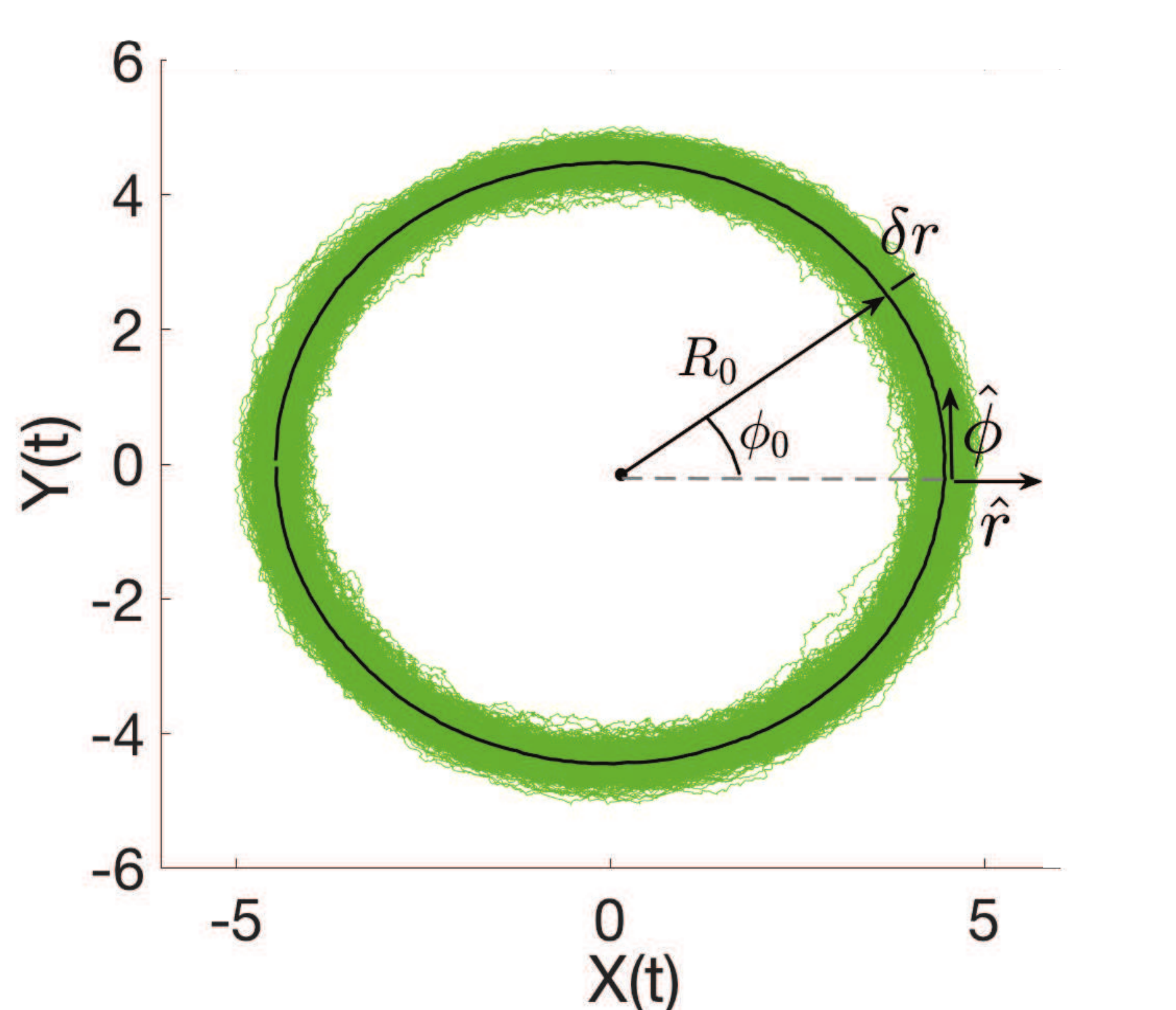}
\colorcaption{A typical stochastic trajectory of a noisy Hopf oscillator based on Eq.~\ref{Hopf-main} is shown in green. Its mean limit cycle is shown as the black circle, which has a 
particularly simple Frenet frame $\{ \hat{r},\hat{\phi} \}$ -- these unit vectors denote the local normal and tangent to the curve, respectively.  
\label{fig:Limit-cycle-sim}}
\end{figure}
We propose that, just as the violation of the original FDT in biological systems is an important quantitative measure of nonequilibrium dynamics~\cite{Gnesotto2018}, the violation of the 
nonequilibrium GFDT provides a quantitative indicator of the presence of an internal active feedback or homeostatic control in biological dynamical systems.  

\section{The stochastic Hopf oscillator}
The simplest dynamical model of hair bundle oscillations is the stochastic supercritical Hopf oscillator in its normal form. This two-dimensional 
dynamical system can be described in terms of a complex variable $z(t) = x(t) + iy(t)$, which obeys the differential equation
\begin{equation}
\label{Hopf-main}
\dot{z} = z \left(  \mu + i \omega \right)  - b z |z|^2 + \eta_z(t) + f_z(t),
\end{equation} 
where $f_z(t)$ is an external deterministic force acting on this overdamped system, and $\eta_z(t)$ is a stochastic force, described below. 
The dynamics of the deterministic and unforced system $(f_z = \eta_z = 0)$ are controlled by the values of the model parameters $\left\{\mu, \omega, b = b' - i b'', (b' , b'' > 0) \right\}$.  
The real parameter $\mu$ 
determines the power input to the system. When  $\mu < 0$, this term damps the oscillations, leaving the system with a single fixed point at $z = 0$, with an infinite basin of attraction. 
As this parameter becomes positive, there is positive energy input into the system, and the oscillator undergoes a supercritical Hopf bifurcation, resulting in a circular limit cycle of radius $R_{0} = \sqrt{\mu/b'}$, which also has an infinite basin of attraction. The oscillator has an angular frequency given by $\omega_{0} = \omega + R_0^{2} b''$, where we assume 
that $\omega$ is real. 

To specify the stochastic system, we include a Gaussian white noise force $\eta_z = \eta_x + i\eta_y$ with a zero mean: 
\begin{eqnarray}
\label{noise-mean}
\langle \eta_{i}(t) \rangle &=& 0,\\
\label{noise-std}
\langle \eta_{i}(t) \eta_{j}(t') \rangle &=& A_{ij} \delta(t-t'),
\end{eqnarray}
with the symmetric $A_{ij}$ ($A_{xy} = 0$) allowing for the uncorrelated noise in the $x$ and $y$ channels to be drawn in principle from different Gaussian distributions.  Finally, we include deterministic external perturbations
via $f_z(t) = f_x(t) + if_y(t)$.

In the following, it will be convenient to work in polar coordinates: $r = \sqrt{x^{2}+ y^{2}}$ and $ \phi = \arctan(y/x)$~\cite{Sheth2019}.
Trajectories derived from Eq.~\ref{Hopf-main} are those of an overdamped particle moving in two dimensions in response to a force field ${\bf f}$, which can be 
decomposed into the gradient of an azithumally symmetric scalar potential energy $\Phi (r)$ and the curl of  a vector potential ${\bf A} = \hat{z} A(r)$, where 
\begin{eqnarray}
\label{scalar-hopf}
\Phi(r) &=& -\frac{\mu}{2} r^{2} + \frac{b'}{4} r^{4},\\
\label{vector-hopf}
A(r) &=& -\frac{\omega}{2}r^{2} -  \frac{b''}{4} r^{4}.
\end{eqnarray}
$\Phi(r)$ is the well-known ``Mexican Hat" potential and is illustrated in Fig.~\ref{fig:Scalar_potential} along with $A(r)$. Also shown is the particle's deterministic, 
limit cycle. The curl of its vector potential $\bf f_v = \nabla \times A(r)$,  is a constant azimuthal force that drives the 
particle circularly along the minima of $\Phi(r)$. In Fig.~\ref{fig:Scalar_potential} we use: $\mu = 40$, $\omega = 10$, $b' = 2$ and $b'' = 2$.

When driven by white noise, the conservative system with $\omega = b'' =0$  corresponds to the case of an overdamped particle in thermal 
equilibrium at some finite temperature.  The vector potential, representing the action of the hair cell's endogenous molecular motors, 
does work on the overdamped system, generating the limit cycle oscillations, as shown in Fig.~\ref{fig:Limit-cycle-sim}. We use the same parameter values as above. Other simulation details are described in Appendix ~\ref{app-sim_details}. 

The appearance of a force field produced by a vector potential does not alone generate a limit cycle or even a nonequilibrium steady state.  The necessary and sufficient conditions to create such a state with a time-independent force field is that: (1) the force field is proportional to the curl of a vector potential, and (2) the force does work on the particle that represents the state of the oscillator.  A classic counterexample, where the second condition is not met, is provided by a charged particle in a magnetic field.  In Appendix ~\ref{app-charge_magnetic_field}, we review this case, showing that a damped, charged particle in a two dimensional harmonic potential and in a uniformly applied magnetic field, aligned in the direction perpendicular to the plane of the charged particle's motion, obeys the standard FDT. 
\begin{figure}
\includegraphics[width=1\linewidth]{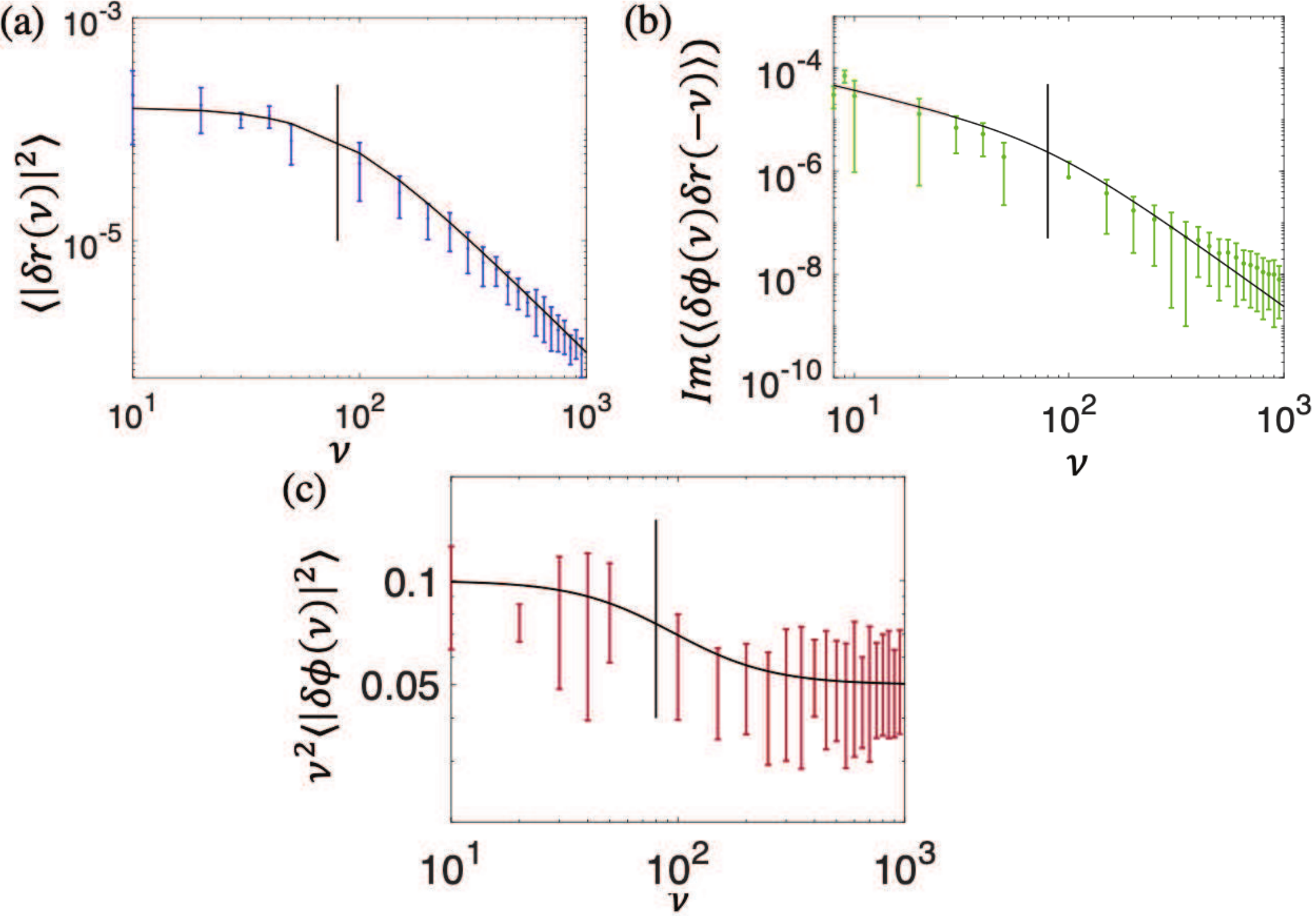}
\colorcaption{Measured two-point correlations (colored points) of fluctuations $\delta r(\nu)$, $\delta \phi(\nu)$	of the simulated stochastic Hopf limit cycle oscillator 
are shown in the frequency domain along with their corresponding analytical calculations (black lines) -- see Eq. \ref{correlation-matrix}. 
Error bars show the standard deviation of the mean. Panel (c) illustrates the frequency-dependent phase diffusion constant. The adaptive 
drive introduces $\delta r$, $\delta \phi$ cross-correlations (panel (b)) so that the radial fluctuations enhance phase diffusion for frequencies below the Lorentzian corner frequency of the radial fluctuations, indicated by the vertical (black) line in all the panels.
\label{fig:correlation-functions}}
\end{figure}

Generally, for hair cell models, one allows the $b$ coefficient to be complex, as mentioned above. In this case, 
the azimuthal drive generates dynamics of the form: $\dot{\phi} = r^{-1} \nabla \times {\bf A} = \omega + b'' r^{2}$.   The power input of the drive is now {\em adaptive}, meaning that 
the work it does on the particle depends on the current state of the system as determined by the radial coordinate.  In the dynamical systems literature, when the 
azimuthal coordinate is driven independently of the state of
the system here given by $r$, \textit{i.e}., when $b''=0$, the system is said to experience isochronous driving. Conversely,  adaptive driving where $b'' \neq 0$ is referred to as nonisochronous. For our purposes, the important feature of this model is that the adaptation of the drive forcing the steady-state limit cycle oscillations can be continuously varied through the 
one model parameter $b''$.  

To study the fluctuations of the system about its limit cycle (when $\mu > 0$), we expand about the limit cycle
\begin{eqnarray}
\label{delta-variable-r}
r(t) &=& R_{0} + \delta r (t) \\
\dot{\phi}(t) &=& \omega_0 + \delta \dot{\phi}(t),
\label{delta-variable-phi}
\end{eqnarray}
to find two coupled stochastic linear Langevin equations for the fluctuations of the radius $\delta r$ and phase $\delta \phi$ of the oscillator
\begin{eqnarray}
\label{dr-noise-dynamics}
\delta \dot{r} &=& -2 \mu \, \delta r + \eta_r + f_r, \\
\delta \dot{\phi} &=& 2  b'' \sqrt{\frac{\mu}{b'}}\,  \delta r  +  \eta_{\phi} + f_{\phi}.
\label{dphi-noise-dynamics}
\end{eqnarray}
Here, the terms $\left\{\eta_r, \eta_\phi\right\}$ and $\left\{f_r, f_\phi\right\}$ are projections of the stochastic and perturbative forces onto the local normal $\hat{r}$ and tangent $\hat{\phi}$, respectively. These unit vectors span the Frenet-Serret frame associated with the averaged limit cycle of the oscillator, being the local normal 
and tangent directions, respectively (see Fig.~\ref{fig:Limit-cycle-sim}). The details of this averaging are in Appendix \ref{app-sim_details}.

The Frenet-Serret frame advances and simultaneously rotates along the mean limit cycle at angular velocity $\omega_0$. By working in this co-moving reference frame, we subtract away the mean non-equilibrium dynamics of the steady-state oscillator. Doing so allows us to recover a GFDT for the nonequilibrium system, as discussed by Seifert and coworkers~\cite{Speck2006}.   Note that the use 
of the dimensionless phase angle $\phi$ instead of the arclength variable $s = R_{0} \phi$ requires the noise amplitudes $\eta_{r,\phi}$ to have different length dimensions.  To account for this explicitly, we set second moments of the Gaussian force fluctuations in the frequency domain (given by $\nu$) to be 
\begin{eqnarray}
\langle \left| \eta_{r}(\nu) \right|^{2} \rangle &=&1\\
\langle \left| \eta_{\phi}(\nu) \right|^{2} \rangle &=&R_{0}^{-2},
\end{eqnarray}
which also has the effect of setting the effective noise temperature to $1/2$, since the mobilities   in the Hopf equation have been set to unity.  To account for this dimensional difference, 
it will be convenient in the following to define a symmetric ``temperature matrix'' by ${\cal T}_{rr} = 1, {\cal T}_{r \phi}  = R_{0}^{-1}, {\cal T}_{\phi \phi}  = R_{0}^{-2}$. 
This choice of coordinates has no other consequences for our analysis.

To verify the GFDT in the co-moving frame, we first compute the correlation matrix in the frequency domain
\begin{equation}
\label{correlation-matrix-template}
\mathbf{{\cal C}}(\nu) =  \begin{bmatrix}
\	\langle |\delta r (\nu) |^2 \rangle& \langle \delta r (\nu) \delta \phi (-\nu)  \rangle\\
\langle \delta \phi (\nu) \delta r (-\nu) \rangle& 	\langle |\delta \phi (\nu) | ^2 \rangle
\end{bmatrix}.
\end{equation}
Using Eqs.\ref{dr-noise-dynamics}, \ref{dphi-noise-dynamics} we obtain 
\begin{widetext}
\begin{equation}
\label{correlation-matrix}
\mathbf{{\cal C}}(\nu) =  \begin{bmatrix}
\frac{1 }{4 \mu^{2} + \nu^2} & 0	\\
0 &     \frac{ b'}{ \mu \nu^2} 
\end{bmatrix} 
+ 2 b''
\begin{bmatrix}
0 & \frac{-i}{\nu (4\mu^2 + \nu^2)}\\
 \frac{i}{\nu (4\mu^2 + \nu^2)} & \frac{2{b''}\mu}{b' \nu^2 \left(4 \mu^{2}+ \nu^2\right)} 
\end{bmatrix}.
\end{equation}
\end{widetext}
The radial autocorrelations are those of an overdamped harmonic oscillator, as expected from the form of the scalar potential in Eq.~\ref{scalar-hopf}, calculated near the circular limit cycle $r = R_{0}$. Similarly, the autocorrelations of the phase angle $\sim \nu^{-2}$, as expected for phase diffusion.  When the drive is nonadaptive or isochronous ($b''=0$), there is a simple, frequency-independent phase diffusion constant, and there are no cross correlations between the radial and phase fluctuations.  The adaptation of the 
drive, however, introduces both a frequency-dependent phase diffusion constant (observed in hair-cell data~\cite{Sheth2018})
and, more importantly, new correlations between the radial and phase fluctuations.  Both of these effects
arise because the internal drive changes its power input in response to the state of system, given by $\delta r$ (Eq.~\ref{dphi-noise-dynamics}).  All three correlation functions are shown in 
Fig.~\ref{fig:correlation-functions}, where the solid (black) lines show the theoretical predictions, and the (colored) points the numerical results from our Brownian simulations.  The 
error bars on the numerical data points represent the standard deviation of the mean. 
\begin{figure}
\includegraphics[width=1\linewidth]{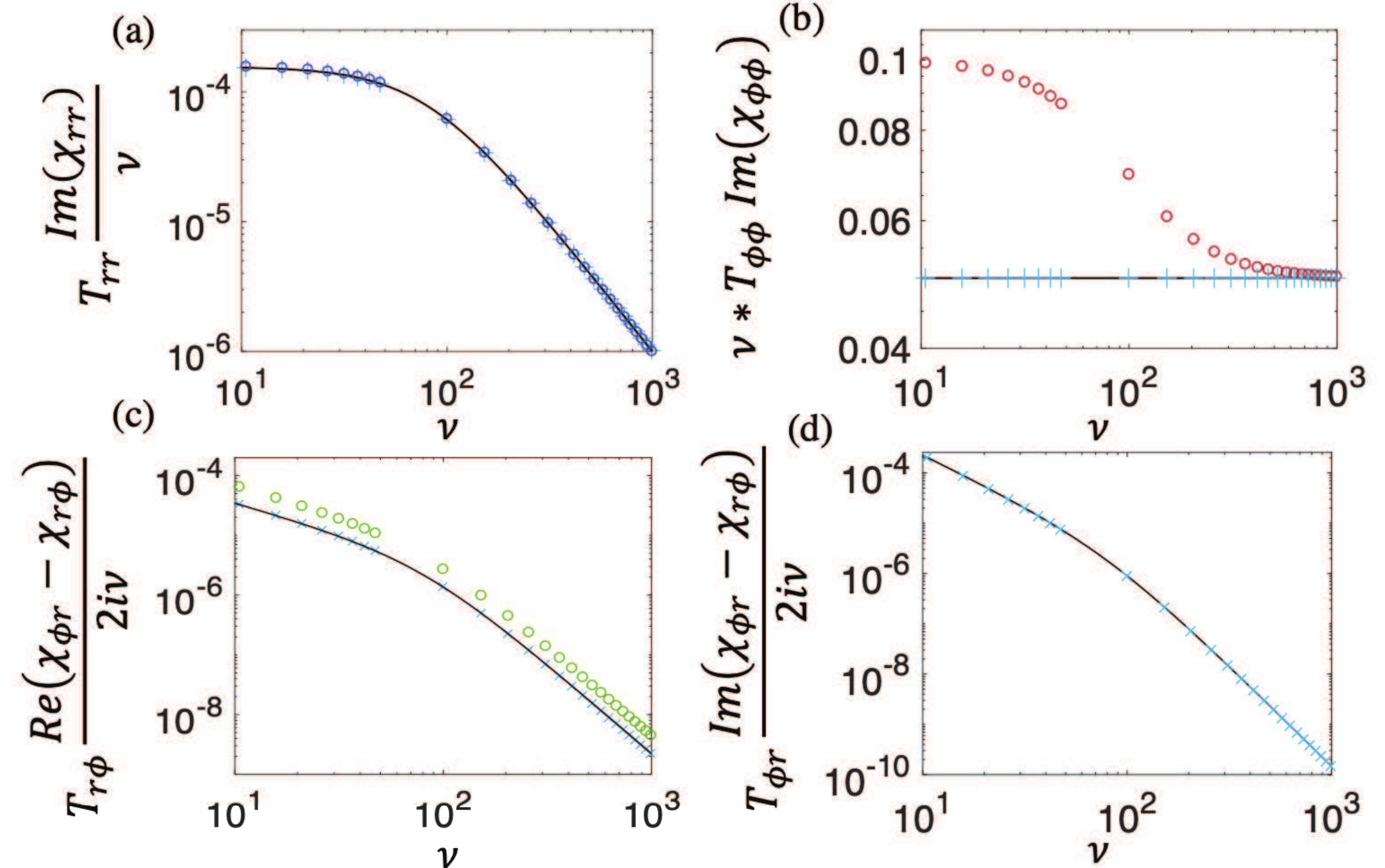}
\colorcaption{Breakdown of GFDT in the adaptively driven Hopf oscillator. We compare the measured two-point correlations of Fig.~\ref{fig:correlation-functions} (circles) with those inferred from numerical response function data via GFDT (light blue crosses). The latter agrees with the analytical calculations (black lines) of $ \left[ \chi_{\alpha \beta}(\nu) - \chi_{\beta \alpha}(-\nu) \right] {\cal T}_{\beta \gamma}$ using Eq.~\ref{response-matrix}. While the GFDT predicted correlation function agrees with observations for the radial fluctuations (panel (a)), it differs from those for the phase diffusion (panel (b)). In the bottom panels, we show the real (panel (c)) and imaginary (panel (d)) parts of $\chi_{\phi r}(\nu) - \chi_{r \phi}(-\nu)$. The former is related to the cross-correlations of $\delta r$ and $\delta \phi$. The GFDT prediction of these correlations also fails (blue crosses vs green circles). 
\label{fig:response-functions}} 
\end{figure}

A direct calculation of the response matrix 
\begin{equation}
x_{\alpha}(\nu) = \chi_{\alpha \gamma}(\nu) f_{\gamma}(\nu)
\end{equation}
gives
\begin{equation}
\mathbf{\chi}(\nu) = 
\begin{bmatrix}
\frac{1 }{2 \mu -i \nu} & 0\\
 -2b''\sqrt{\frac{\mu}{b'}}\frac{1}{(i\nu) (2\mu - i\nu)}& -\frac{1}{i \nu} 
\end{bmatrix}.
\label{response-matrix}
\end{equation}
We define the deviation from the GFDT matrix as 
\begin{equation}
\label{delta-gfdt}
\Delta_{\alpha \beta}(\nu) = \left[ \chi_{\alpha \beta}(\nu) - \chi_{\beta \alpha}(-\nu) \right] {\cal T}_{\beta \gamma} - 2 i \nu {\cal C}_{\alpha \gamma}(\nu),
\end{equation}
and find that deviations from the GFDT (which is the FDT in the co-moving frame associated with the mean limit cycle) appear only in the presence of an adaptive drive, \textit{i.e.} when $b'' \neq 0$:
\begin{equation}
\label{GFDT-deviations}
\Delta(\nu) = 2b''
\begin{bmatrix}
0 &  \frac{(-\nu + 2 i\mu)}{(4\mu^2 \nu  + \nu^3)}\\
 \frac{(\nu + 2 i\mu)}{(4\mu^2 \nu  + \nu^3)} & \frac{-i 4{b''}\mu}{b' \nu \left(4 \mu^{2}+ \nu^2\right)}
\end{bmatrix}.
\end{equation}
Only the Lorentzian fluctuations of the radial $\delta r$ variable obey the GFDT when the drive is adaptive.  When $b'' \neq 0$, 
the feedback between the azimuthal driving force and the radial oscillations
breaks the GFDT due to both new cross correlations  $C_{r \phi}$ and the modified phase diffusion seen in $C_{\phi \phi}$.  
This breakdown of the GDFT cannot be removed by an appropriate change of
variables, as has been explored for nonequilibrium fluctuations about a fixed point~\cite{Prost09}. 

In Fig.~\ref{fig:response-functions}(a), we show the correspondence between the correlation data obtained from numerical 
simulations (dark blue circles) and that expected from the response function (light blue crosses) for the radial variable based on the GFDT. 
In Fig.~\ref{fig:response-functions}(b), where we compare the frequency-dependent phase diffusion constant measured from the numerical data (red circles) and the 
GFDT-based prediction (blue crosses), we see the failure of the GFDT for the adaptively driven system. 
Clear deviations are seen at low frequencies, as predicted by Eq.~\ref{GFDT-deviations}. When the drive is 
not adaptive ($b'' = 0$) -- see Fig.~\ref{fig:hopf_isochronous} in appendix~\ref{app-no_adaptation} -- these deviations vanish. The GFDT is once again obeyed. 
We also show the real and imaginary parts of $\chi_{\phi r}(\nu) - \chi_{r \phi}(-\nu)$ in panels (c) and (d) respectively. The former predicts the 
cross-correlations of the radial and phase fluctuations via GFDT. Those predicted blue crosses illustrated in panel (c) also fail to agree with the 
simulation data (green circles). In all panels (a - d), we show our analytical calculations of $ \left[ \chi_{\alpha \beta}(\nu) - \chi_{\beta \alpha}(-\nu) \right] {\cal T}_{\beta \gamma}$ as obtained from Eq.~\ref{response-matrix}  (black lines). These are in universal agreement with the GFDT-based 
correlation functions inferred from numerically simulated response function data (light blue crosses).

\section{Adaptively driven three-state model}

\begin{figure}
	\includegraphics[width=1\linewidth]{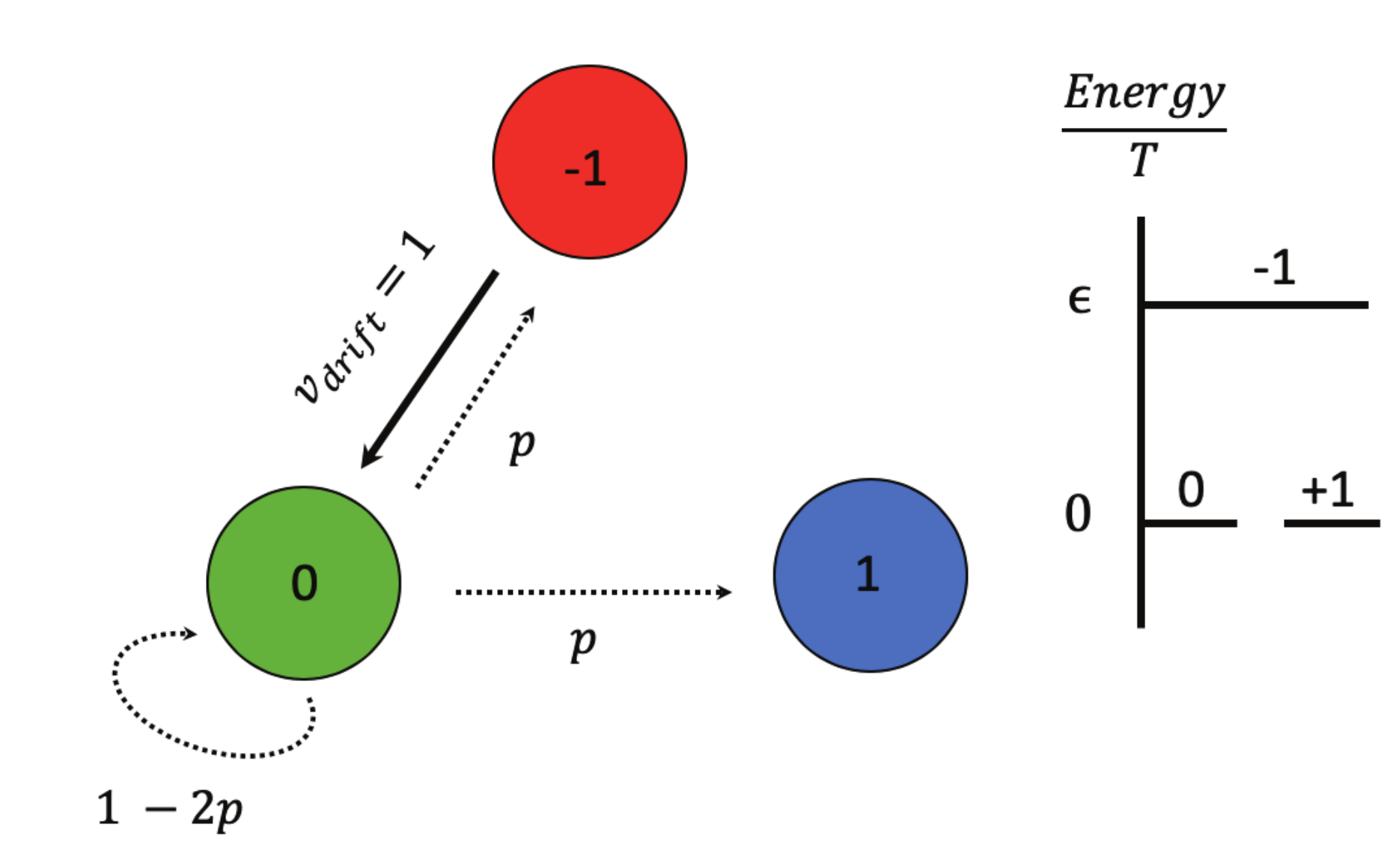}
	\colorcaption{A schematic diagram of the three-state system, showing states $\{-1,0,1\}$ denoted by red, green, and blue disks respectively. These states have 
		energies $\{\epsilon, 0,0 \}$.  In the equilibrium steady state, $v_{\rm{drift}} = 0$. Conversely, when $v_{drift} = 1 $ or 2,  the system has a non-zero internal drive. The 
		resulting steady state probability current may be removed by working in a co-moving frame. 
		\label{fig:3_state_system_model_a}}
\end{figure}

To better understand the role of adaptive feedback on the drive in breaking GFDT, it is helpful to examine the same phenomenon in a more simple, finite-state model. We analyze two such three-state systems. First, as shown 
in Fig.~\ref{fig:3_state_system_model_a}, we consider a system with three states labeled by $s = \{-1,0,+1\}$ and having energies $\{\epsilon, 0, 0\}$. 
The system's discrete-time dynamics combine a drift velocity $v_{\rm{drift}} = 0,1,2$ anticlockwise around the triangle of 
states -- see Fig.~\ref{fig:3_state_system_model_a} -- and stochastic hopping. 
The $v_{\rm{drift}}$ plays the role of the drive, breaking detailed balance in the system. The hopping rate $p$,  $0 \leq p \leq 0.5$, is 
unbiased when $\epsilon = 0$ and generates diffusion amongst the three states. 
For example, the deterministic system (p = 0) with $v_{\rm{drift}} = 1 (2)$ generates uniform anticlockwise (clockwise) motion of state occupation 
around the triangle shown in Fig.~\ref{fig:3_state_system_model_a}. Meanwhile the stochastic system ($p > 0$) with 
no drift ($v_{\rm{drift}} = 0$) obeys detailed balance and corresponds to an equilibrium system. The role of the constant drift is to introduce a 
finite state analog of the nonadaptively driven Hopf model ($\omega_{0} \neq 0, b" = 0$). We will later incorporate an adaptive drive by 
allowing the value of $v_{\rm{drift}}$ to temporally depend upon the history of state occupation which allows us to study 
the finite-state analog of the adaptively-driven Hopf oscillator ($\omega_{0} = 0, b" \neq 0$).

We incorporate adaptation by setting
\begin{equation}
\label{adaptive-drift-velocity}
v_{drift} (t_i) =  \lfloor \sum_{j = 1}^{\infty} r e^{\lambda \left(i - j\right)} \xi \left(t_{i-j}\right) \rfloor \mod 3,
\end{equation}
where $\lfloor. \rfloor$ is the floor function returning the integral part of its argument. We have also introduced the function $\xi(t_{i-j})$, which takes the value $1,-1$ or $0$ when the system is in state $1,-1$ or $0$ respectively at time $t_{i-j}$. In turn $\xi(t_{i})$ is defined using the indicator functions $\sigma_k(t_{i})$ which are 1 (0) when the system is (is not) in state $k$ at time $t_{i}$:
\begin{equation}
\label{xi-definition}
\xi\left(t_{i} \right) = \left( -1\right)^{\sigma_{-1} (t_i)} \left[ 1 - \sigma_{0} (t_i)\right].
\end{equation}
Finally we note that the adaptive drive depends on two constants $r$, which determines the responsiveness of the adaptation, and $\lambda$, which sets the memory time ~$\lambda^{-1}$ for the drive.

\begin{figure}
	\includegraphics[width=1\linewidth]{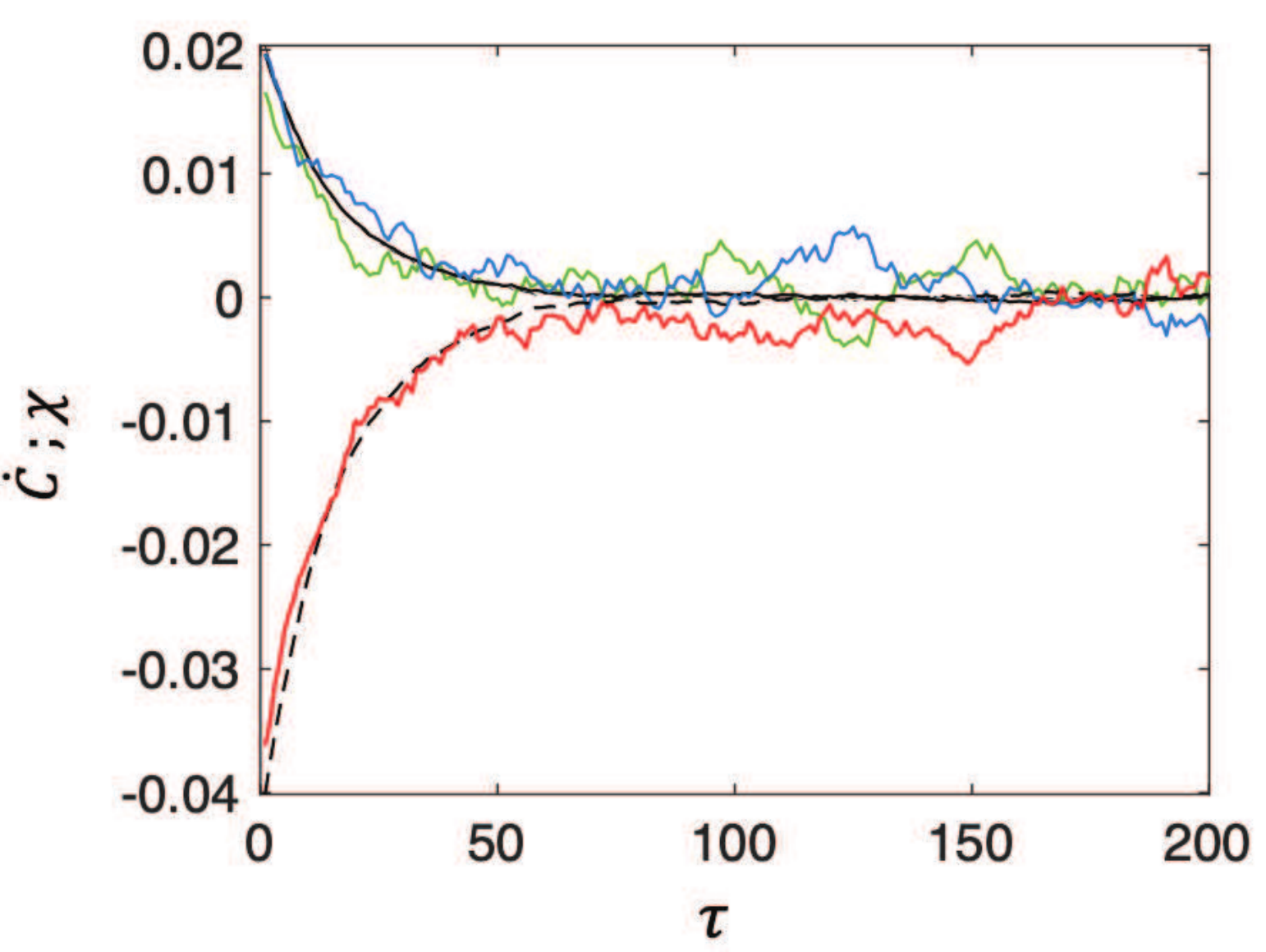}
	\colorcaption{Test of FDT for the equilibrium system.  By comparing $\dot{C}_{-1,-1}$ (black dashed line) and $\chi_{-1,-1}$ (red line), we check that the 
		response of the system to a force driving it out of the $-1$ state matches the appropriate correlation function derivative.   We also find the expected 
		correspondence between $\dot{C}_{0,-1}$ (black solid line) and $\chi_{0,-1}$ (green line) as well as $\dot{C}_{1,-1}$ (black dashed-dot line) and  $\chi_{1,-1}$ (blue line). 
		\label{fig:only_diffusion_all_states}}
\end{figure}

We first perform numerical simulations of the symmetric model ($\epsilon =0$) that obeys detailed balance ($v_{\rm{drift}} = 0$). We tracked the stochastic trajectories of 40 realizations of the system 
over a total of $4 \times 10^4$ time steps for each of the realizations. For additional details of these simulations, we refer the reader to Appendix ~\ref{app-sim_details}. 
Setting $\epsilon = 0$ resulted in the occupation probability of the three states being one third, as expected (not shown).  From these
trajectories, we also compute all two-point correlation functions 
\begin{equation}
\label{symmetrized-correlation}
C_{n m}(\tau) = \frac{1}{2} \left[ \langle \sigma_{n}(t_{i} + \tau ) \sigma_{m}(t_{i}) \rangle  + \langle \sigma_{m}(t_{i} + \tau ) \sigma_{n}(t_{i}) \rangle\right].
\end{equation}
The average is taken over an ensemble of trajectories at  time delay $\tau$.  Under the assumption of ergodicity, one may alternatively average over longer time series from one trajectory.  
Further, an experimentalist investigating the stochastic dynamics of a nonequilibrium steady state system might implicitly assume time reversal invariance. Therefore our definition of the correlation function was chosen to make it explicitly time reversal invariant when $n \neq m$.  Clearly, if the driven system admits a non-vanishing 
probability current, this symmetry will not be valid. However, since we propose using the violation of fluctuation theorems as a test for both an underlying limit cycle in general 
and an adaptively driven one in particular, we will suppose {\em a priori} that the correlation data is analyzed assuming time-reversal invariance in the steady state. 

To  test the standard FDT, we numerically obtained the response of the occupation probability of state $n$, $p_n(t) = \langle \sigma_n (t) \rangle$, to a force conjugate to the occupation of state $-1$,
\begin{equation}
\label{response-function-def}
\delta p_{n}(t_{i}) = - \sum_{j=-\infty}^{i}\chi_{n,-1}(t_{i} - t_{j})\delta \epsilon_{-1}(t_{j}),
\end{equation}
by setting the energy of that 
state to $\epsilon_{-1}=3$ for one time step and observing the subsequent stochastic evolution of the system. We confirmed that our perturbation was in the linear response regime by varying $\epsilon_{-1}$ -- see appendix \ref{app-linear}.
In Fig.~\ref{fig:only_diffusion_all_states}, we plot $\chi_{-1, -1}(\tau)$, $\chi_{0, -1}(\tau)$, and $\chi_{1, -1}(\tau)$ as solid red, green, and blue 
curves respectively. As expected, the transient increase in the energy of the $-1$ state suppresses the occupation probability of that state and symmetrically increases the occupation probability
of the other two states: $+1$ and $0$.  The system recovers its equilibrium probabilities exponentially with a decay rate of about 20 inverse time units. 

The standard FDT requires that these response functions must be equal to the time derivative of the correlation functions $\dot{C}_{n,-1} (\tau)$ evaluated at time delay $\tau$.  We plot the 
numerically obtained time derivatives of the correlation functions $\dot{C}_{-1, -1}$, $\dot{C}_{0, -1}$ and $\dot{C}_{1, -1}$ as dashed black, solid black, and dashed-dot black lines respectively in Fig.~\ref{fig:only_diffusion_all_states}.  As expected, we find that the time derivatives 
of the correlation functions of state occupation agree with the responses of the occupation probability to a force conjugate to that variable. All the correlation functions in this figure were normalized such that $C_{n,m}(0) = \delta_{n,m}$. The response functions were multiplied by an empirical temperature, in this case 0.2. Remaining computations for this system are all similarly normalized.

\begin{figure}
	\includegraphics[width=1\linewidth]{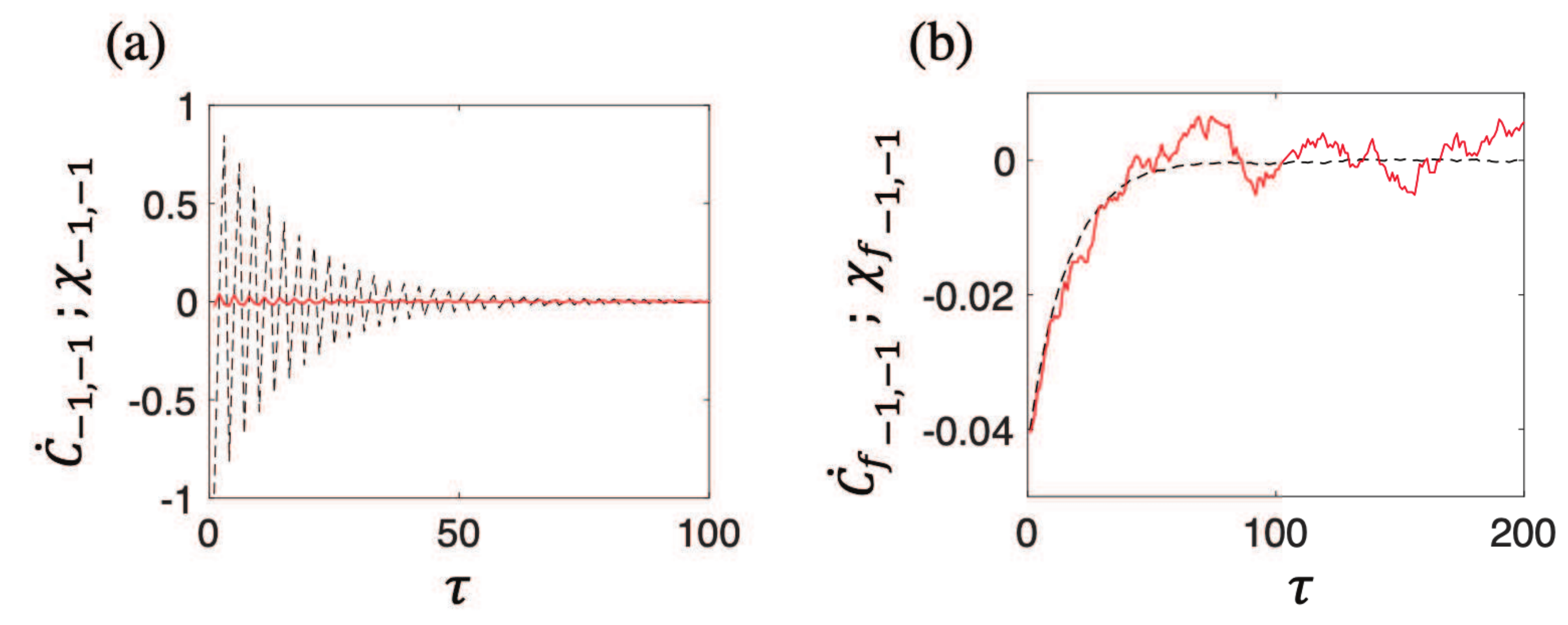}
	\colorcaption{(a) In the presence of an internal drive, the system violates FDT, as can be seen in the comparison of its linear response function $\chi_{-1,-1}$ (red) and the time derivative of its auto-correlation function $\dot{C}_{-1,-1}$ (black dashed). (b) However, upon transforming to the co-moving reference frame, we show that the three-state model satisfies GFDT. The derivative $\dot{C}_{f-1,-1}$ (dashed black) and the response function $\chi_{f-1,-1}$ (red) are now in agreement with each other.
	\label{fig:constant_drift_diffusion_state-1}}
\end{figure}

We now consider the case of a constant drive, setting $v_{\rm{drift}} = 1$. In Fig.~\ref{fig:constant_drift_diffusion_state-1}(a), we demonstrate the violation of FDT by this system. The red curve is the numerically computed response function $\chi_{-1,-1} (\tau)$, and the dashed black curve is the derivative of the corresponding correlation function $\dot{C}_{-1,-1}(\tau)$, whose oscillatory nature can be attributed to the internal drive of the model. The standard FDT requires these to be equal. They are not, indicating breakdown of FDT. However, for this model we propose that one may obtain a valid GFDT by evaluating the correlation and response functions in a reference frame co-moving with velocity $v_{\rm{drift}}$. To transform to the co-moving frame, we introduce new indicator functions,
\begin{equation}
\label{tilde-sigma-definition}
\tilde{\sigma}_i\left(t_{j} \right) =  \sigma_{(i + v_{\rm_{drift}} t_j) \mod 3}.
\end{equation}
We find that in the co-moving frame the numerically computed response function of $\chi_{f -1,-1}$ (red curve) agrees with $\dot{C}_{f -1,-1}$ (dashed black line) as seen in Fig.~\ref{fig:constant_drift_diffusion_state-1}(b). The values of these two functions, as a matter of fact, are similar to those of the equilibrium system (Fig.~\ref{fig:only_diffusion_all_states}). Due to the symmetry of the problem, we only show plots for the -1 state.  For the other two states the reader is referred to Appendix~\ref{app-no_adaptation}. 
\begin{figure}
	\includegraphics[width=1\linewidth]{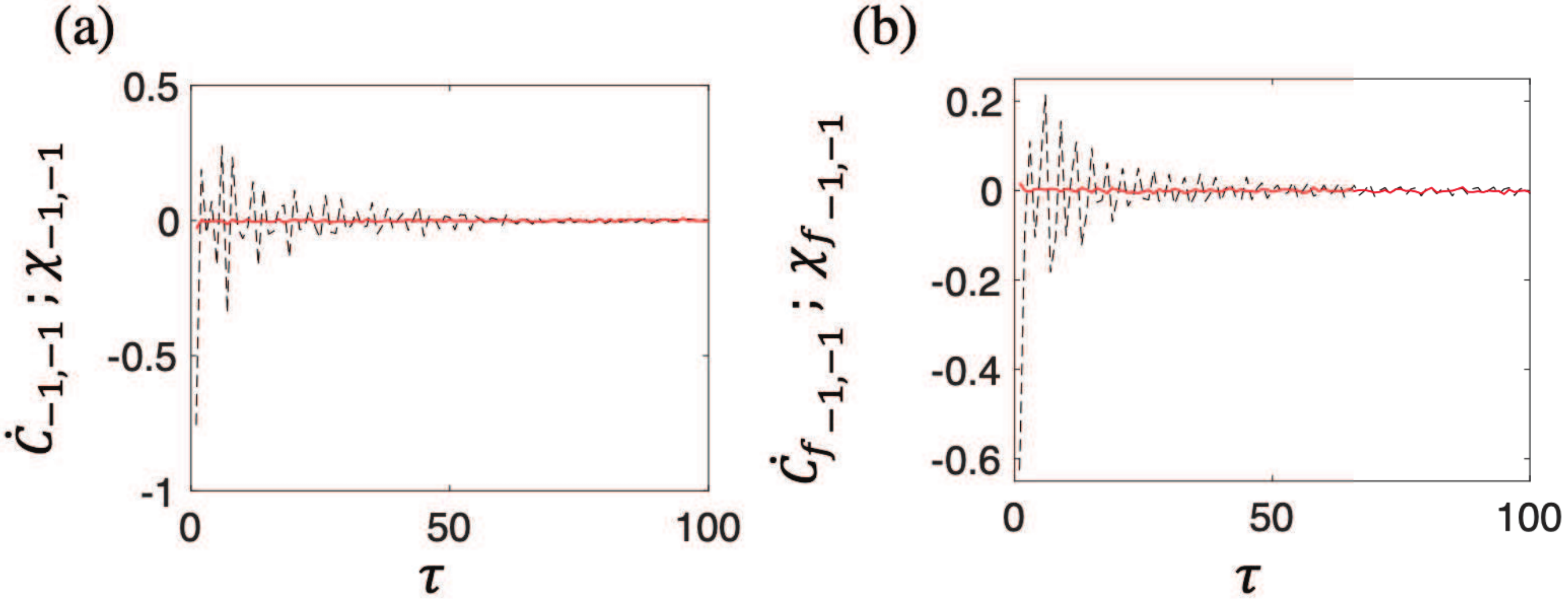}
	\colorcaption{The three-state model with an adaptive drive (see Eq.~\ref{adaptive-drift-velocity}). (a) We juxtapose the derivative $\dot{C}_{-1,-1}$ (dashed black) and the response function $\chi_{-1,-1}$ (red) to illustrate the breakdown of FDT. (b) Furthermore, unlike the one illustrated in Fig.~\ref{fig:constant_drift_diffusion_state-1}, this system also violates GFDT, as is evident by comparing $\dot{C}_{f -1,-1}$ and $\chi_{f -1,-1}$, computed in the associated co-moving frame. 
		\label{fig:history_drift_diffusion_state_-1}
	}
\end{figure}

Next, we study the adaptively driven three-state model choosing $r = 2$ and $\lambda = 0.1$. This non-Markovian system violates FDT as shown 
by the plots in Fig.~\ref{fig:history_drift_diffusion_state_-1}(a). The time derivative of the auto-correlation $C_{-1,-1}$ (black dashed line) deviates 
appreciably from the response function $\chi_{-1,-1}$ (shown in red). Plots for the other two states are given in Appendix~\ref{app-adaptation-three-state}. 
Moreover, in the co-moving frame, the adaptation of the internal drive precludes restoration of the generalized theorem (Fig.~\ref{fig:history_drift_diffusion_state_-1}(b)). 
In order to test the GFDT in the rotating frame we chose a reference frame co-moving with the average drift velocity, which in our simulations was 1. There exists 
no other reference frame that may restore the GFDT in the adaptively driven system. The breakdown of both FDT and GFDT relations in this system is 
similar to the adaptively-driven Hopf limit cycle with parameter $b" > 0$.

One may ask whether any time variation of the drive is sufficient to invalidate the FDT or the GFDT.  To address this, we considered a randomly varying drive that has the 
same average drift velocity as the one examined above.  We consider the three-state model with a randomly varying drift velocity that has equal probabilities at each time step of being 
0,1,or 2. There are no temporal correlations in the stochastic $v_{\rm drift}$. It is easy to see that the mean drift velocity is unity.   
This system, unlike the adaptively-driven three-state model,
obeys the GFDT (data not shown).  We conclude that state-dependent feedback on the drive is required to invalidate the GFDT.

 \begin{figure}
\includegraphics[width=1\linewidth]{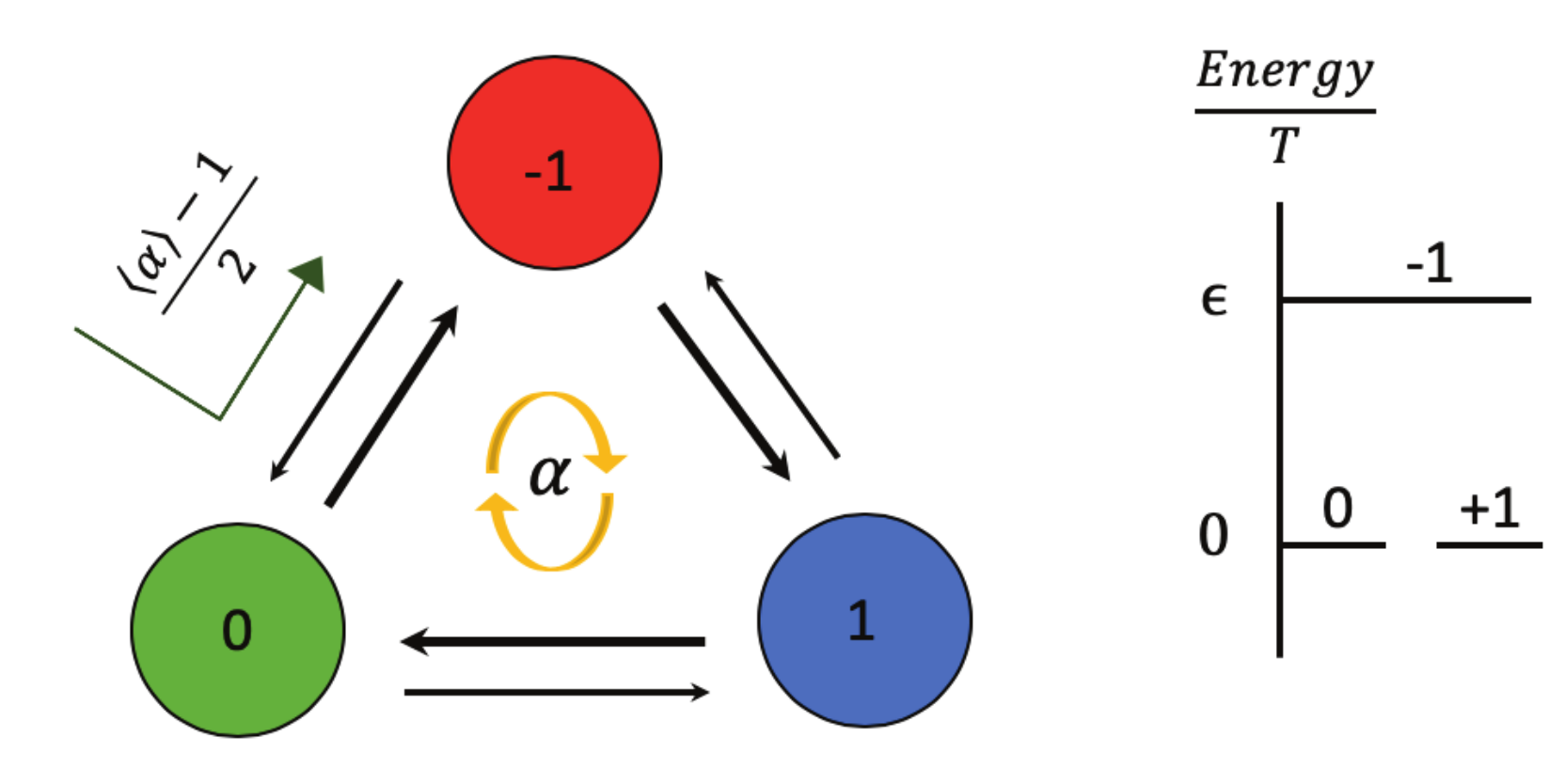}
\colorcaption{A schematic diagram of the three-state system, showing states $\{-1,0,1\}$ denoted by red, green, and blue disks respectively. These states have 
energies $\{\epsilon, 0,0 \}$.  In the nonequilibrium steady state, the clockwise transitions rates are enhanced over their detailed-balance values by $\alpha(t)$.  The 
resulting steady state probability current may  again be removed by working in a co-moving frame. 
\label{fig:3_state_system_model_b}}
\end{figure}

We also examine the stochastic dynamics of a more general three-state model defined by the discrete-time master equation for the probability $p_n(t)$ of observing the system in state n = -1,0,1 at time $t_i$ :
\begin{equation}
\label{master-equation}
 p_n \left(t_{i+1} \right) = \sum\limits_{m \neq n} \left[p_m(t_{i}) \alpha_{mn}(t_{i}) - p_n(t_{i}) \alpha_{nm}(t_i) \right].
\end{equation}
See Fig.~\ref{fig:3_state_system_model_b} for a schematic representation. The system evolves via six transition probabilities, {\em e.g.}, the transition rate from state $n$ to $m$ at time $t_{i}$:  $\alpha_{nm}(t_{i})$. 
These six transition probabilities are given by the following rules.  We set 
\begin{equation}
\label{detailed-balance}
\alpha_{nm}  = \alpha_{mn} e^{\epsilon_n - \epsilon_m} \alpha(t_{i}),
\end{equation}
where $n$ is to the right of $m$ in the list of states $\{-1,0,1\}$ or its cyclic permutations.  The factor $\alpha(t)$ allows us to drive the system into a 
nonequilibrium steady-state by breaking detailed balance.  By choosing $\alpha(t)$ to be a constant greater than one, we generate a clockwise probability current -- see Fig.~\ref{fig:3_state_system_model_b} -- 
in steady state.  Such a choice is analogous to turning on a non-adaptive drive in the isochronous Hopf model ($\omega_{0} \neq 0, b'' = 0$) 
of the hair cell oscillator.  Later, to introduce an adaptive process, we will consider the 
case in which the drive depends upon the history of the system by setting 
\begin{equation}
\label{drive-memory}
\alpha\left(t_{i} \right) = 1 + r \sum_{j = 1}^{\infty}  e^{\lambda \left(i - j\right)} \xi \left(t_{i-j}\right),
\end{equation}
where $\xi(t_{i-j})$ has been defined in Eq.~\ref{xi-definition}.

The strength of the adaptation is again controlled by $r$. $\lambda$ controls the exponential decay rate of the 
memory kernel in Eq.~\ref{drive-memory}.  It is measured in inverse time units $\delta t = t_{i+1} - t_{i}$, which we always set to 0.01.   
The effect of the feedback is to increase the drive when the system 
has recently been in the $+1$ state and decrease it when the system has visited the $-1$ state.  The simulations for this three-state model, as for the one before, were performed using 40 realizations over $4\times10^4$ time steps. 

\begin{figure}
\includegraphics[width=0.9\linewidth]{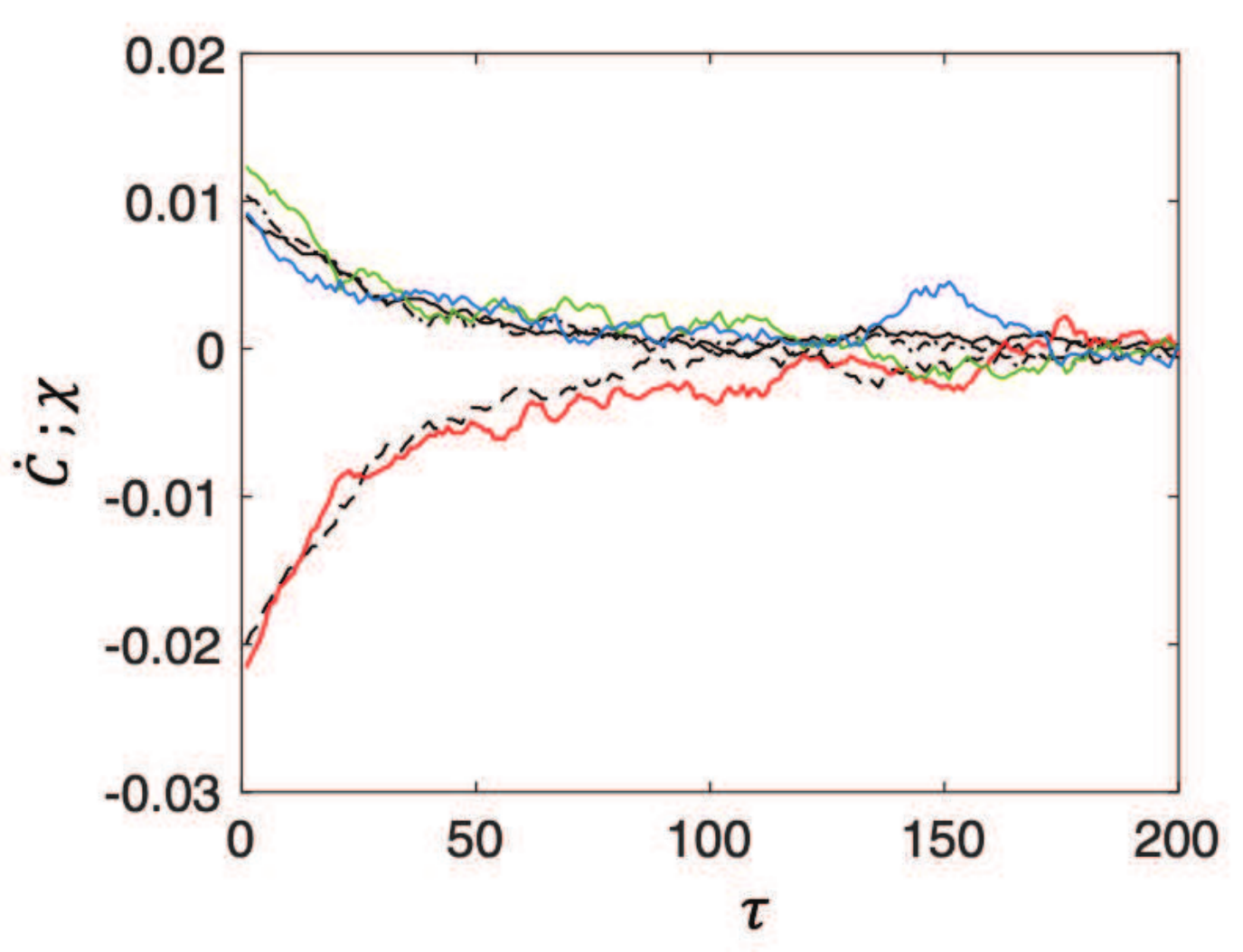}
\colorcaption{The FDT relation is satisfied by the detailed-balance system. As in  Fig.~\ref{fig:only_diffusion_all_states}, we compare the response of the system to a change in the energy of state -1 with the corresponding correlation function derivatives. We show $\dot{C}_{-1,-,1}$ (dashed black), $\dot{C}_{0,-,1}$ (solid black), $\dot{C}_{1,-,1}$ (dot dashed black), $\chi_{-1,-,1}$ (red), $\chi_{0,-,1}$ (green) and $\chi_{1,-,1}$ (blue). 
\label{fig:no_history_alpha_1}}
\end{figure}

We first study the detailed-balance system, which can easily be shown to be equivalent to the non-driven case of the first three-state model. In Fig.~\ref{fig:no_history_alpha_1}, 
we illustrate the time derivatives $\dot{C}_{-1,-1} (\tau)$, $\dot{C}_{0,-1} (\tau)$ and $\dot{C}_{1,-1} (\tau)$ as the dashed black, solid black and dot-dashed black lines respectively. 
Also shown are the linear response functions of $\chi_{-1,-1} (\tau)$ (red), $\chi_{0,-1} (\tau)$(green) and $\chi_{1,-1} (\tau)$(blue), which, as anticipated, overlap with their 
corresponding correlation derivatives. The empirical temperature of this system is $0.11$.

\begin{figure}
\includegraphics[width=1\linewidth]{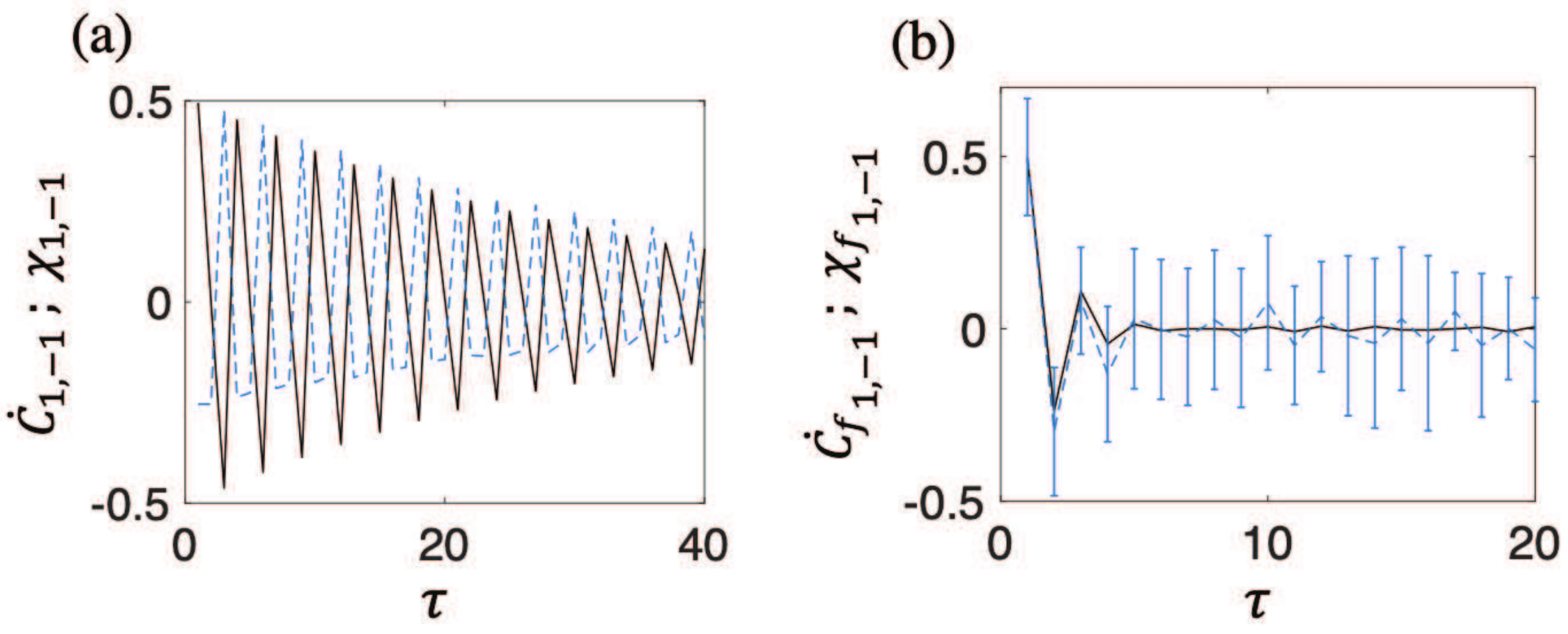}
\colorcaption{(a) FDT violation in the broken detailed balance system at $\alpha(t) = 98$. We compare the time derivative of the 
cross-correlation $\dot{C}_{1, -1}$ (black line) and the response function $\chi_{1, -1}$ (blue dashed line).  (b) In the co-moving frame the 
GFDT holds as seen by comparing correlation $\dot{C}_{f_{1, -1}}$ (black line) and response function $\chi_{f_{1, -1}}$ (dashed blue line). 
Error bars denote the standard deviation of the mean. 
\label{fig:no_history_alpha_98}}
\end{figure}

We now repeat this measurement in a non-equilibrium system by setting $\alpha(t) = 98$.
This choice of a non-adaptive drive breaks detailed balance and is similar to
the Hopf model of hair cell oscillations with $b''=0$ but $\omega>0$.   In Fig.~\ref{fig:no_history_alpha_98}(a), we show the measured response function $\chi_{1, -1}(\tau)$.  As in the equilibrium case, the applied force pushes the system out of the $-1$ state into the $0, 1$ states.  But, unlike the equilibrium case, 
the change in probability oscillates in time due to the detailed-balance-breaking drive.  For example, the occupation probability of $+1$ cycles the three-state system in the clockwise direction while
slowly decaying over longer times (not shown), resulting in an oscillatory response function as in the dashed blue line in the figure.  The correlation 
function $C_{1,-1}(\tau)$ also shows this oscillatory behavior, but its derivative (black line) {\em does not} match the corresponding response 
function. The standard FDT is violated.  

We can, however, obtain a GFDT in the driven system by working in a ``rotating'' reference frame -- one that moves with the clockwise 
probability current of the non-equilibrium steady state.   Unlike the previous three-state model the Frenet frame state occupation variables are not well-defined. The frame's velocity is determined by the mean probability current of the system. Moving at the speed of $\frac{\alpha - 1}{2}$,
we now find that the response function of the $+1$ state in this co-rotating frame to a force acting on the $-1$ state -- the dashed blue line in Fig.~\ref{fig:no_history_alpha_98}(b) -- agrees with 
the numerically measured time derivative of the correlation function (calculated using Eq.~\ref{symmetrized-correlation}), shown as the black line in this figure.  The error bars represent the standard deviation of the mean for the response
data.   We find a similar agreement between the other correlation and response functions in the co-rotating frame; these are shown in Appendix ~\ref{app-no_adaptation} (Figs.~\ref{fig:no_history_alpha_98_neg1} and~\ref{fig:no_history_alpha_98_0}). 
While neither the time derivative of the correlation function nor the response function in the driven system agrees with predictions based on the equilibrium system, their agreement with 
each other shows that a generalized fluctuation dissipation theorem holds in the driven system, as expected based on the work of the Seifert and collaborators~\cite{Seifert2010}.  The appearance of 
the GFDT in the co-rotating frame, which zeros out the steady-state probability current of the driven system, is analogous to our observation of a similar fluctuation theorem in the 
driven Hopf oscillator system without the adaption.  

\begin{figure}
\includegraphics[width=1\linewidth]{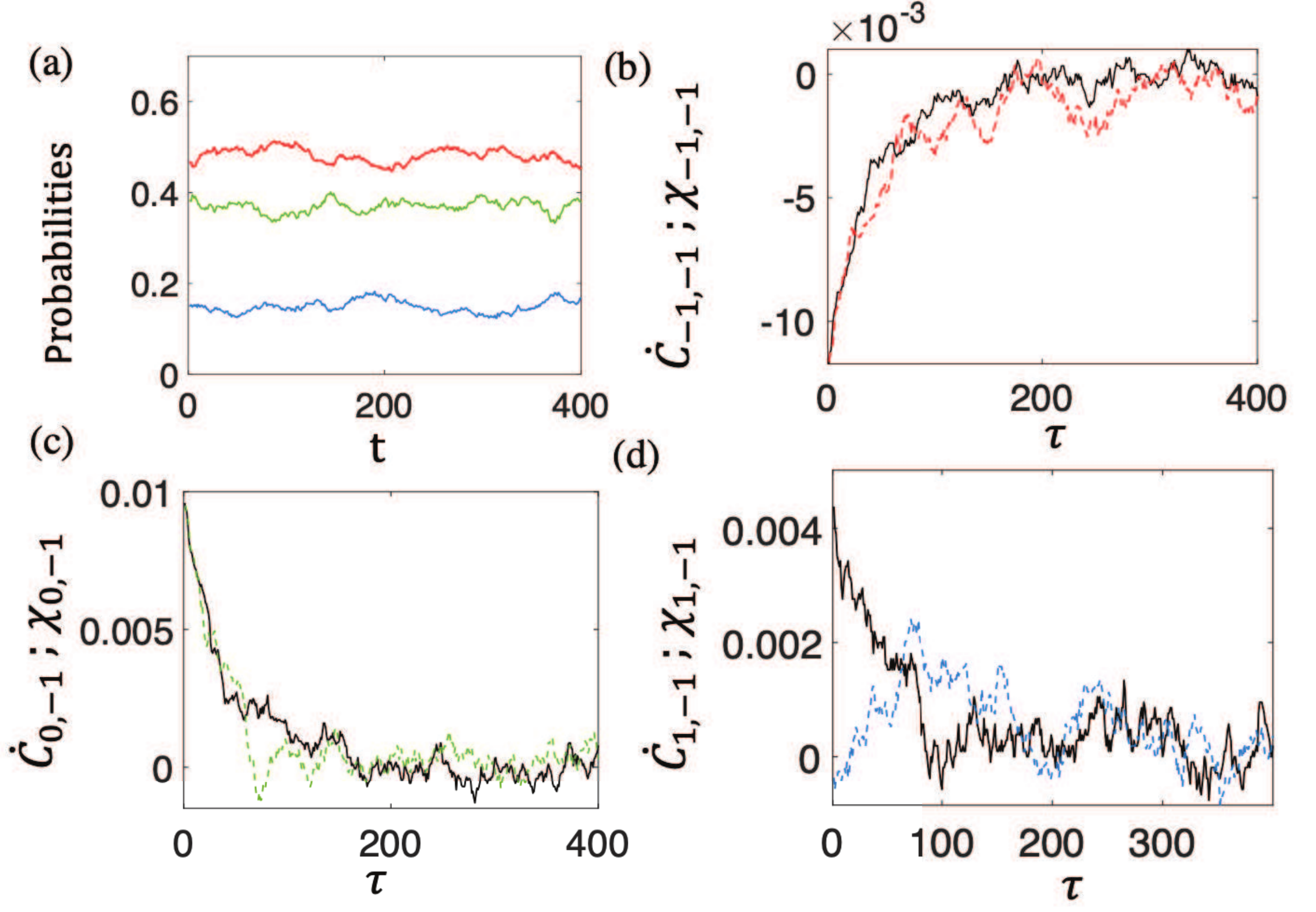}
\colorcaption{The three-state system with an adaptive drive, Eq.~\ref{drive-memory}.  (a) With the adaptive drive at r = 0.095 and $\lambda =0.1$, the states are no longer occupied with
equal probability even when their occupation energies are equal.  In panels (b)-(d) we compare $\dot{C}$ (black solid) with the appropriate $\chi$ (color) in the ground frame.  
In (d) we observe significant deviations from the FDT.  
\label{fig:history_alpha}}
\end{figure}

We now reinstate an adaptive drive in this more general three-state system via Eq.~\ref{drive-memory}, taking $r = 0.095$ and $\lambda = 0.1$. This is analogous to the adaptively driven Hopf system. We obtain a steady-state system with non-equal occupation probabilities of the three states in steady state.  In spite of the fact that the energies of all three states are equal, the 
adaptive drive breaks the permutation symmetry of these states, as shown in panel (a) of Fig.~\ref{fig:history_alpha}.  As a result, the simple occupation probabilities of the states in this 
nonequilibrium steady state do not reflect their relative energies.  Conversely, just by observing these occupation probabilities, one might conclude erroneously that this system was in equilibrium with a particular spectrum of energy levels.  To test this conclusion, one must not only examine these probabilities but also compare the correlation and response functions of the system. 

In the three remaining panels (b-d) of Fig.~\ref{fig:history_alpha}, we show a comparison of the time derivative of the correlation function $\dot{C}_{k, -1}(\tau)$ and the 
response function $\chi_{k,-1}(\tau)$ for $k = -1, 0, +1$ in panels (b), (c), and (d) respectively. The standard FDT fails vividly for one set of measurements:  $\dot{C}_{1,-1}(\tau) \neq \chi_{1,-1}(\tau)$ -- see Fig.~\ref{fig:history_alpha}(d). 
\begin{figure}
	\includegraphics[width=1\linewidth]{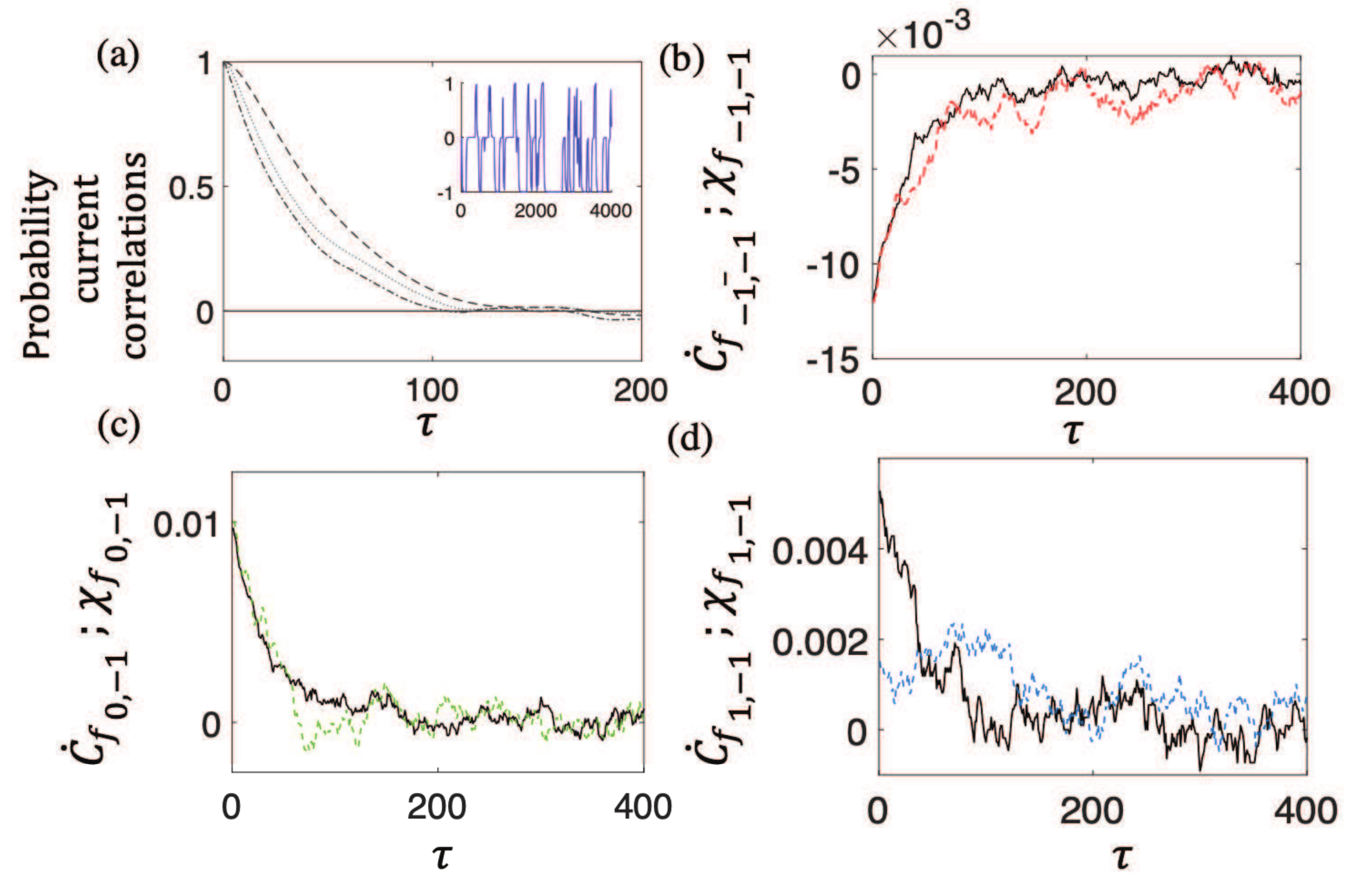}
	\colorcaption{(a) Correlation of the steady-state probability current $\alpha(t) -1$ for different $\lambda$ and $r$ values: $r = 0.095, \lambda = 0.1$ (dashed line),  $r = 0.095, \lambda = 0.25$ (dotted line),  $r = 0.095, \lambda = 1$ (dot-dashed line) and  $r = 0$ (solid). For $r = 0.095$ and $\lambda = 0.1$ , we show the stochastic current from a representative trajectory (inset). In panels (b)-(d), we illustrate the violation of the GFDT when working in a co-moving frame that works to eliminate the mean probability current. Comparing with Fig.~\ref{fig:history_alpha}, we see that this frame partially, but rather imperfectly restores the fluctuation theorem. 
	\label{fig:history_alpha_frenet}}
\end{figure}

We examine the probability current in this more general three-state model.  We show the temporal correlation function of the current in the adaptively driven system 
in Fig.~\ref{fig:history_alpha_frenet} (a). In the inset of the same panel we show a representative part of the time series of the probability current from which the correlation functions 
were obtained.   Clearly, as $\lambda$ is decreased (drive memory time increased) the probability current's correlation time increases, so that the drive becomes less adaptive. Its current value
depends on a long time average of the system, which itself necessarily varies only slowly in time.  As a result, we find that with sufficiently long memory times, the adaptively driven system begins
to resemble a nonadaptively driven one, so long as $r/\lambda$ remains fixed.  As a result, the magnitude of the violations of the GFDT will decrease.

Given this intuition, it is interesting to examine the residual violation of the GFDT in a system driven by a weakly adaptive drive.   
Due to the current fluctuations that are still correlated with the state of the system, it is clear that no co-moving frame can precisely reestablish the 
GFDT. But we can find the best approximation to the GFDT in this system by working in a co-moving reference frame selected to eliminate the 
mean probability current, {\em i.e.}, we chose a velocity  $\frac{\langle \alpha  \rangle - 1}{2}$ to minimize GFDT discrepancies.  
In the remaining panels of Fig.~\ref{fig:history_alpha_frenet} (b-d) we do this.
The results shown in Figs.~\ref{fig:history_alpha_frenet}(c) and \ref{fig:history_alpha_frenet}(d) demonstrate that the GFDT still fails 
due to feedback between the system and the drive. But a comparison between Fig. \ref{fig:history_alpha_frenet}(c) and 
Fig. \ref{fig:history_alpha_frenet}(d) measured in the co-moving frame with 
Fig. \ref{fig:history_alpha}(c) and Fig. \ref{fig:history_alpha}(d), showing the same quantities in the non-rotating lab frame, demonstrates the partial restoration of the GFDT. 
With even weaker drive adaptation, this restoration of the GFDT further improves (data not shown).

\section{Summary}

Systems that exhibit nonequilibrium steady states frequently violate the FDT. Failure to satisfy the conditions set by that theorem has therefore been used as a test of the 
nonequilibrium nature of various stochastic steady-states, indicating the presence of an energy consuming process. As complex biological systems invariably 
contain active processes, such a test is useful for experimentally quantifying precisely which degrees of freedom in the system are out of equilibrium and thereby 
learning something about the underlying processes maintaining that nonequilibrium steady state. For example, in actomyosin gels, one 
observes enhanced strain fluctuations at low frequencies due to 
motor activity. This is a consequence of the fact that the motor dynamics introduce force autocorrelations with a colored noise spectrum.  
As a result, the strain fluctuations observed across multiple timescales do not correspond to the equilibrium (visco-)elastic system at any single temperature.  
In this and other systems, the quantitative measurement of the breakdown of the FDT is a type of sensor for the detection of nonequilibrium steady states and 
for providing a measure of how far from equilibrium they are. 

Broadly speaking, there are multiple ways to violate the FDT, and, as a consequence, not all FDT violations have the same implications.  Another class of nonequilibrium 
biological systems break detailed balance and have non-vanishing probability currents in steady state. We have explored one particular class of such systems: those exhibiting 
stochastic limit cycles.  In this case, previous work has introduced a new class of 
generalized fluctuation theorems, GFDTs, based on working in a co-moving reference frame that effectively eliminates the stationary probability current. When 
fluctuations are viewed in this co-moving reference frame, the familiar relations between them and the response functions of the system are 
restored.

In the current work, we introduce a new feature: the internal drive maintaining the limit cycle, in effect, measures the state of the system and adapts its power input 
based on that measurement. In systems driven by such an adaptive drive, we demonstrated the violation of the GFDT.  We first examined the 
Hopf oscillator model with an adaptive drive, where the quantitative degree of GFDT violation is 
proportional to single model parameter $b''$, which measures the ability of the azimuthal drive to adapt its power input in response to the 
radial excursions of the system.  

To isolate the role of drive adaptation in breaking the GFDT in an even simpler model, we introduced two related three-state systems defined by 
discrete time master equations.  In these we violate detailed balance
by producing stationary states with a nonzero probability current. These systems violate the standard FDT, as expected.  By 
introducing a co-rotating frame to eliminate the probability current in the three-state system without an adaptive drive
 we obtain a GFDT, which is consistent with previous work.  But when we incorporate drive
adaptation by allowing the probability current to change based on the history of the system's trajectory, we once again observe the breakdown of the GFDT. We also observed
that the feedback between the drive and the state of the system is crucial for GFDT violations.  These vanish if the drive's power input varies randomly in time in a manner
uncorrelated with the state of the system. 
These results are in direct
analogy with the more complex Hopf model and allow us to more carefully probe the role of drive adaptation in the violation of the GFDT. 

The Hopf model is, in fact, the simplest model for describing the dynamics of hair cell motion. As such, it 
provides an important connection between the basic questions of fluctuation theorem theorems (or their 
failure) in adaptively driven steady states and stochastic dynamics in a living system.   It also presents us with a relatively simple biological 
dynamical system in which to experimentally explore fluctuation theorems in nonequilibrium steady states. Our previous work looking at hair cell 
fluctuations in a Frenet frame that is co-moving with the mean probability current of the system generated correlation data consistent with the theory discussed here~\cite{Sheth2018}.  Future
work is needed to examine the response functions of the system in order to test the GFDT.  

Based on our current work, we propose that, just as the failure of the FDT has been used to test for nonequilibrium steady states, one should be 
able to look for the breakdown of the GFDT as a test of stochastic steady states driven out of equilibrium by an adaptive drive. Two emblematic features of living systems are long-lived nonequilibrium steady states and homeostasis.  One method to maintain homeostatic control of driven states is through an adaptive drive, as seen in the non-isochronous hair 
cell model. As many biological systems may contain homeostatic control that is not as readily accessible experimentally, we suggest that the breakdown of the GFDT may 
serve as a useful tool to indicate the presence of  and to quantify the efficacy of such feedback-based control.  One experimentally tractable biological system in which one 
might test the correlation and causation between an internal adaptive drive and violation of GFDT is the spontaneously oscillating hair cell of the inner ear.

\acknowledgements
AJL acknowledges partial support from NSF-DMR-1709785.  DB acknowledges partial support from 
NSF Physics of Living Systems, under grant 1705139. JS acknowledges partial support from the Fletcher Jones Foundation fellowship.

\clearpage
\newpage

\appendix

\section{Simulation details}
\label{app-sim_details}
\subsection{Hopf oscillator}
The stochastic and externally perturbed Hopf oscillator of Eq.~\ref{Hopf-main} was simulated using the $4^{th}$- order Runge-Kutta method for a duration of 60s, with a time step of  $10^{-4}$s. We explore a large range in the amplitude of the noise variances $\langle \eta_x^2 \rangle$ and $\langle \eta_y^2 \rangle$ (where, $\langle \eta_x^2 \rangle$ = $\langle \eta_y^2 \rangle$) covering $10^{-7}$ to  $0.4$, as well as a range of perturbative forces $10^{-3}$ to $10^{-1}$. All throughout, the amplitude of mean limit cycle oscillators was 
held to be $O(1)$. While consistent results were obtained over the full span of these values, Figs.~\ref{fig:correlation-functions} and~\ref{fig:response-functions} 
employ the highest value of force and noise in their respective ranges. 

\subsection{Mean limit cycle of the Hopf oscillator}
The Hopf oscillator's phase space $\{- \pi, \pi \}$ is partitioned into nearly 200 bins. Trajectories in each bin are then averaged, resulting in the mean curve. 

\subsection{Three-state model}
Eq.~\ref{master-equation} was numerically computed using a random number generator that outputs a value in the range [0 - 1]. Comparison of this value with the 
occupation probabilities of the three states determines the stochastic trajectory for each of the  40 realizations. 
Further, since we define Fig.~\ref{fig:3_state_system_model_b} in terms of transition rates, these probabilities are the product of the respective rates and a time 
step duration of $10^{-2}$.  Data was always taken after 
running the system long enough so that its initial conditions were no longer relevant.
All simulations were performed in MATLAB (R2019a, the MathWorks, Natick, MA).

\section{Electrically-charged particle in a magnetic field}
\label{app-charge_magnetic_field}
The motion of a damped, harmonically bound charged particle of mass $m$ and charge $e$  in the $xy$ plane under the influence of magnetic field $H \hat{z}$ is given by,
\begin{equation}
\ddot{\hat{r}} +  \gamma\dot{\hat{r}} + \omega_0^2 \hat{r} = \frac{e}{mc} \dot{\hat{r}} \times H 
\end{equation}
where $\gamma$ is the friction coefficient, $\omega_0$ is the natural frequency of the oscillator ($\omega_{0} = \sqrt{k/m}$ for a Hookean spring constant $k$), 
and $c$ the speed of light. 
The equations of motion may be written in terms of $x$ and $y$ as
\begin{eqnarray}
\label{x-charged-particle}
\ddot{x} + \gamma\dot{x} + \omega_0^2 x &=& \frac{eH}{mc} \dot{y}, \\
\label{y-charged-particle}
\ddot{y} + \gamma\dot{y} + \omega_0^2 y &=& -\frac{eH}{mc} \dot{x}
\end{eqnarray}
with introduction of the classical Larmor frequency $\omega_r  = \frac{eH}{mc}$. Upon driving Eqs.~\ref{x-charged-particle} and~\ref{y-charged-particle} using either stochastic or deterministic 
(externally applied) forces,  we obtain:
\begin{widetext}
\begin{equation}
\label{mag_noise}
\begin{bmatrix} x\\ y \end{bmatrix} 
= \frac{1}{(-\omega^2 + \omega_0^2 - i\omega\gamma)^2 - \omega_r^2 \omega^2} \times  \begin{bmatrix}
-\omega^2 + \omega_0^2 - i\omega\gamma & -i\omega_r \omega \\ 
i\omega_r \omega & -\omega^2 + \omega_0^2 - i\omega\gamma
\end{bmatrix}
\begin{bmatrix}
\eta_x \\ \eta_y
\end{bmatrix}
\end{equation}
\end{widetext}
When considering these as Langevin equations, we assume rotationally symmetric thermal noise so that $\langle\eta_x^2\rangle = \langle\eta_y^2\rangle = \langle\eta^2\rangle$.

Since the dynamics in directions $\hat{x}$ and $\hat{y}$ are symmetric, we compute and compare one of each of the autocorrelation and cross-correlation functions.  A lengthy but 
straightforward calculation yields the following response and correlation functions.  In order to confirm the validity of the FDT, we present the response functions in combinations such 
that these combinations should be equivalent to the corresponding correlation functions.  We find: 
\begin{widetext}
\begin{eqnarray}
\label{chi_xx}
\frac{\tilde{\chi}_{xx}(\omega) - \tilde{\chi}_{xx}(-\omega)}{2 i } &=& \frac{\gamma\omega((\omega_0^2 - \omega^2)^2 + \gamma^2\omega^2 + \omega_r^2\omega^2)}{(\omega^2\gamma^2 + (\omega_0^2 - \omega^2 - \omega \omega_r)^2)(\omega^2\gamma^2 + (\omega_0^2 - \omega^2 + \omega \omega_r)^2)}\\
\label{c_xx}
C_{xx} &=&  \frac{\langle\eta^2\rangle((\omega_0^2 - \omega^2)^2 + \gamma^2\omega^2 + \omega_r^2\omega^2)}{(\omega^2\gamma^2 + (\omega_0^2 - \omega^2 - \omega \omega_r)^2)(\omega^2\gamma^2 + (\omega_0^2 - \omega^2 + \omega \omega_r)^2)}
\end{eqnarray}
\begin{eqnarray}
\frac{\tilde{\chi}_{xy}(\omega) - \tilde{\chi}_{yx}(-\omega)}{2 i }   &=& \frac{ i\omega_r\omega (\omega_r^2 \omega^2 - (-\omega^2 + \omega_0^2 + i \omega \gamma)^2 + (-\omega^2 + \omega_0^2 - i \omega \gamma)^2 - \omega_r^2 \omega^2)} {2 (\omega^2\gamma^2 + (\omega_0^2 - \omega^2 - \omega \omega_r)^2)(\omega^2\gamma^2 + (\omega_0^2 - \omega^2 + \omega \omega_r)^2)} \\
\label{chi_xy}
&=& \frac{\gamma \omega(2i\omega^3 \omega_r - 2i\omega_0^2 \omega \omega_r^2)}{(\omega^2\gamma^2 + (\omega_0^2 - \omega^2 - \omega \omega_r)^2)(\omega^2\gamma^2 + (\omega_0^2 - \omega^2 + \omega \omega_r)^2)}\\
\label{c_xy}
C_{xy} &=& \frac{\langle\eta^2\rangle(2i\omega^3\omega_r - 2i\omega\omega_r\omega_0^2)}{(\omega^2\gamma^2 + (\omega_0^2 - \omega^2 - \omega \omega_r)^2)(\omega^2\gamma^2 + (\omega_0^2 - \omega^2 + \omega \omega_r)^2)}
\end{eqnarray}
	\end{widetext}
By direct comparison of Eqs.~\ref{chi_xx}, \ref{c_xx},  as well as the cross correlations Eqs.~\ref{chi_xy} and \ref{c_xy}, we verify that FDT is satisfied for a system responding to a 
magnetic field. Even though the force is generated from the curl of a vector potential (like our driving force in the Hopf system), the magnetic field does not invalidate the FDT since the
magnetic forces cannot do work on the system. 

\section{Driven systems without adaptation}
\label{app-no_adaptation}

\begin{figure}
	\includegraphics[width=1\linewidth]{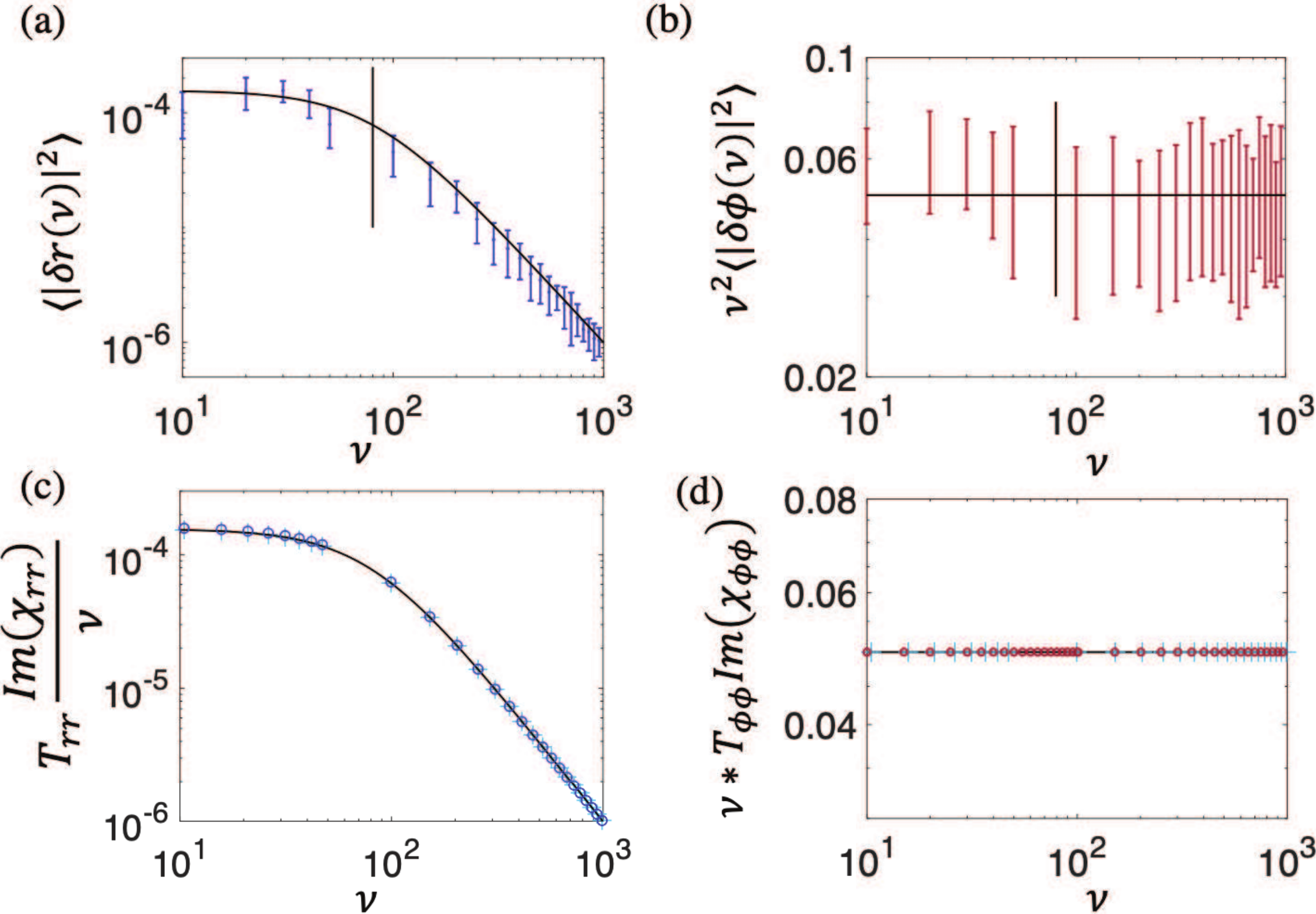}
	\colorcaption{Correlation and response functions for the isochronous Hopf oscillator (b" = 0 ).  (a) Power spectral density of radial fluctuations as a function of frequency $\nu$ (blue dots). (b) Phase diffusion constant, obtained from the product of the phase fluctuation power spectral density and $\nu^{2}$ (red dots). In both panels, the vertical (black) line indicates the corner frequency of $2\mu$. In panels (c) and (d) we compare the measured two-point auto correlation functions with those inferred via GFDT from the numerically obtained response function data of $\chi_{rr}$ and $\chi_{\phi \phi}$. The predicted correlation functions agree with those directly measured from the Hopf oscillator simulations for both the radial and phase fluctuations. Overlaid on all four plots are the respective theoretical calculations from Eqs.~\ref{correlation-matrix} and \ref{response-matrix}.
		\label{fig:hopf_isochronous}}
\end{figure}

In the main text, we present three representative systems that incorporate an adaptive drive. For completeness, we show results obtained from the Hopf system with a nonadaptive 
drive, {\em i.e.}, one with $b''=0$.  This system without an adaptive drive admits a GFDT.   
In Fig.~\ref{fig:hopf_isochronous} we show that the response (black lines) and fluctuations (colored dots) agree as expected from 
the GFDT, or the FDT in the Frenet frame that is co-moving with the mean probability current in the driven oscillator.  
The fluctuations in the normal (radial) direction (blue) are still well described by a simple Lorentzian (black), whose corner frequency is once again marked by a vertical line. 
However, the phase diffusion constant exhibits no frequency dependence (red) consistent with Eq.~\ref{correlation-matrix}. Furthermore, the 
cross correlations $C_{r \phi}$ vanish, and the correlation data depicted in subplots (c and d) agree with those inferred from GFDT and the numerically computed response functions. When $b" = 0$, the hair cell 
model violates FDT but obeys GFDT.

\begin{figure}
	\includegraphics[width=1\linewidth]{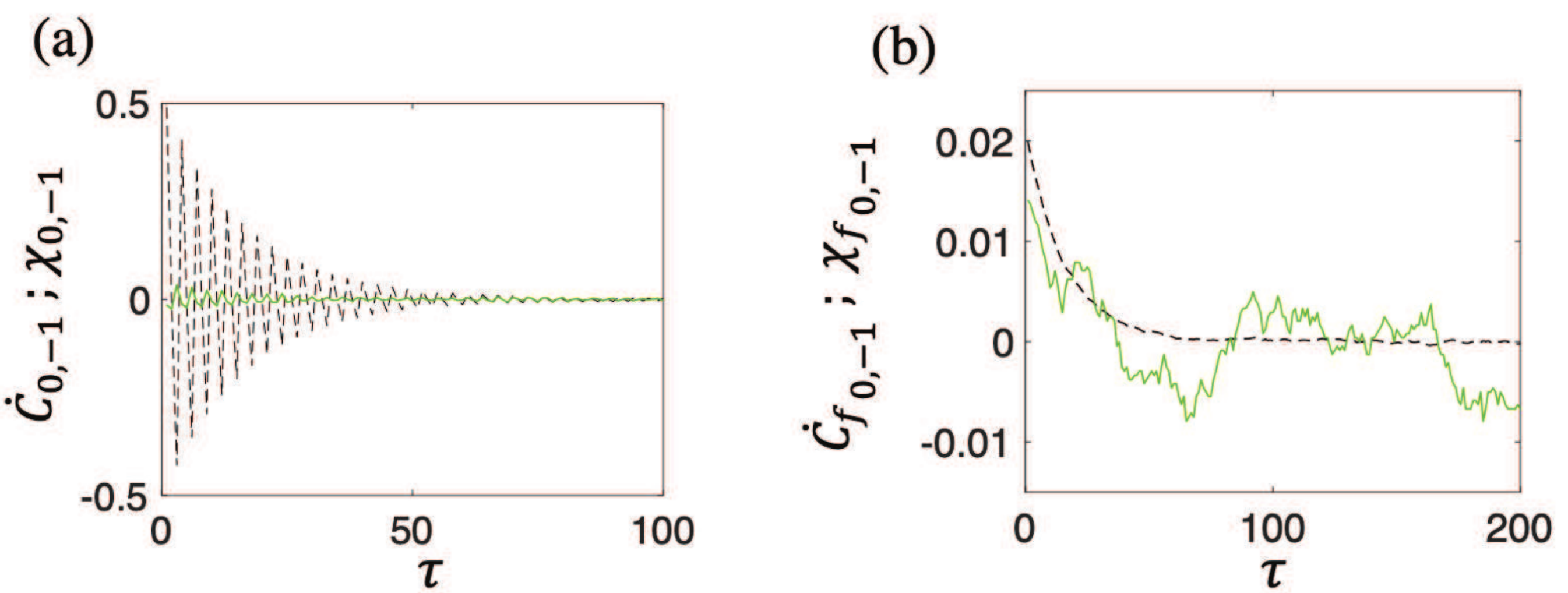}
	\colorcaption{$\dot{C}_{0,-1}$ (black dashed) vs $\chi (\tau)_{0,-1}$ (green solid) for $v_{\rm{drift}} = 1$. (a) Time derivative of the numerically computed cross-correlation function $\dot{C}_{0,-1}$ and the linear response function of $\chi_{0,-1}$ disagree revealing the breakdown of FDT. (b)  However the system satisfies GFDT as seen on comparing these functions calculated in the Frenet frame. 
		\label{fig:constant_drift_diffusion_state_0}}
\end{figure}

\begin{figure}
	\includegraphics[width=1\linewidth]{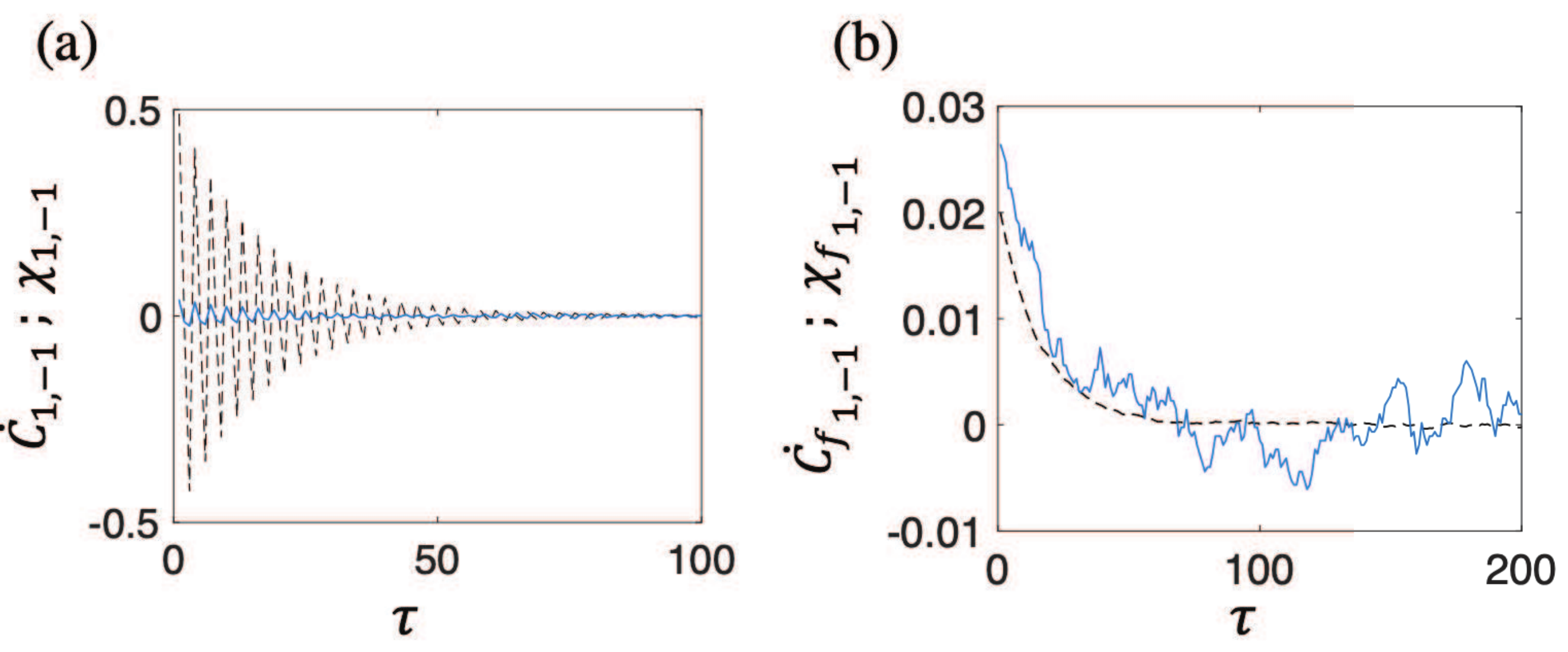}
	\colorcaption{$\dot{C}_{1,-1}$ (black dashed) vs $\chi (\tau)_{1,-1}$ (blue solid) for $v_{\rm{drif}t} = 1$. Comparing the time derivative of the cross-correlation $\dot{C}_{1,-1}$ and response function $\chi_{1 -1}$ we observe significant deviations from FDT in (a) and the satisfaction of GFDT in (b).
		\label{fig:constant_drift_diffusion_state_1}}
\end{figure}

For the stochastic three-state system with a constant $v_{\rm{drift}}$ (Fig.~\ref{fig:3_state_system_model_a}), we have shown in the main text for state -1 that the FDT breaks down, but the GFDT is satisfied. In Figs.~\ref{fig:constant_drift_diffusion_state_0} and \ref{fig:constant_drift_diffusion_state_1} we illustrate the same for the other two states, where we obtain similar results. 

\begin{figure}
	\includegraphics[width=1\linewidth]{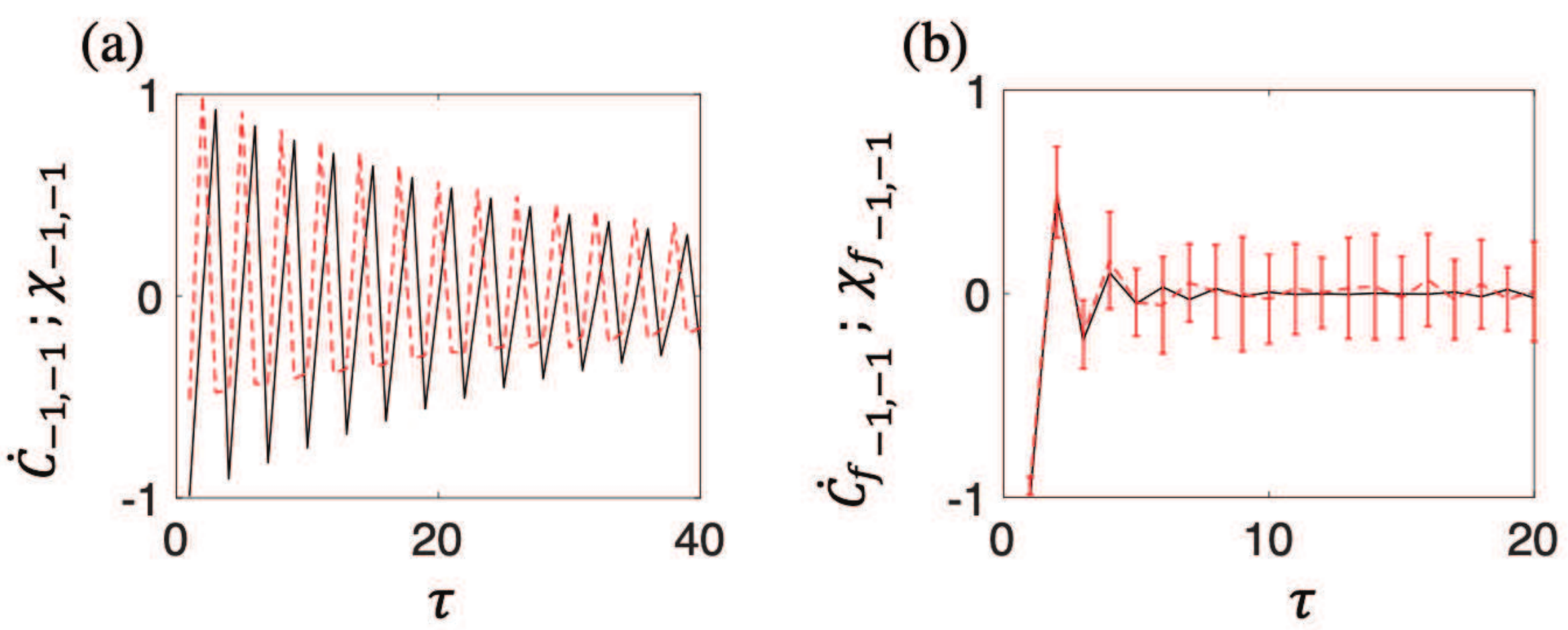}
	\colorcaption{$\dot{C}$ (black solid) vs $\chi (\tau)$ (red dashed) for $\alpha = 98$ and state -1. (a) Time derivative of the cross-correlation $\dot{C}_{-1 -1}$ and the response functions $\chi_{-1 -1}$ superimposed illustrate the breakdown of FDT.  (b)  The Frenet frame formalism allows for the obedience of GFDT. 
		\label{fig:no_history_alpha_98_neg1}}
\end{figure}

\begin{figure}
	\includegraphics[width=1\linewidth]{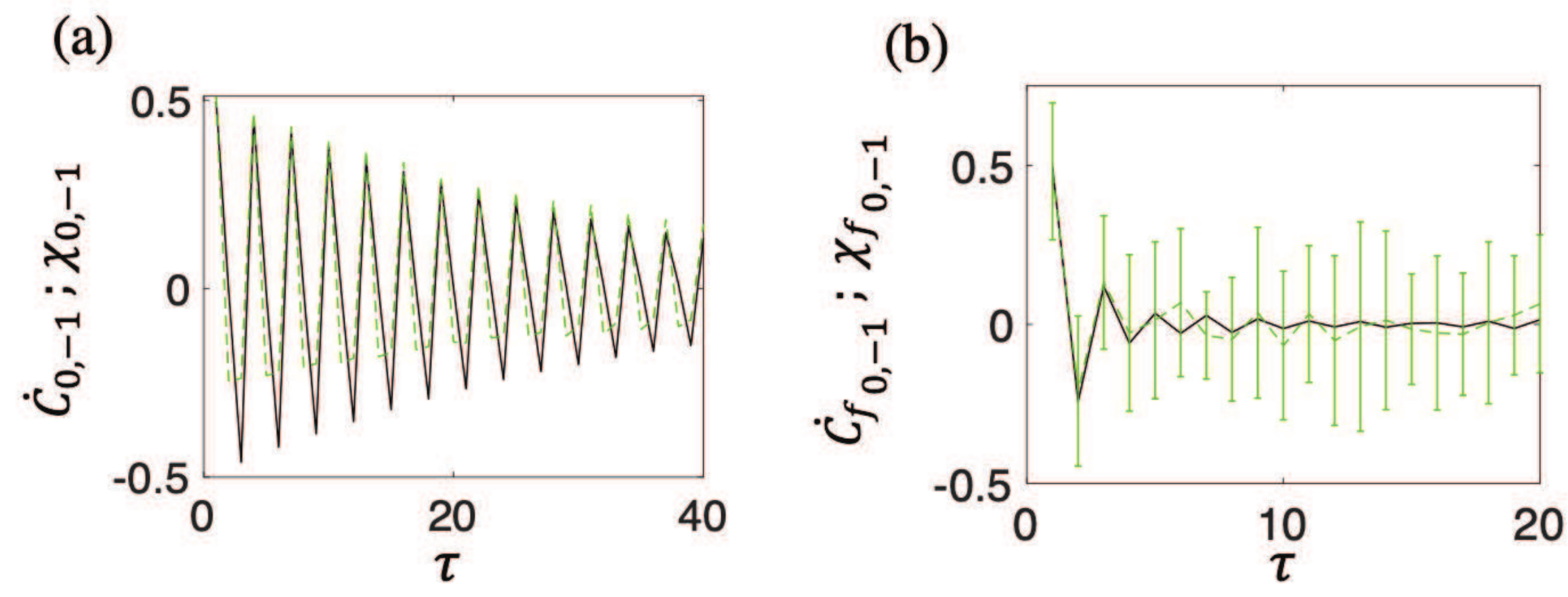}
	\colorcaption{$\dot{C}$ (black solid) vs $\chi (\tau)$ (green dashed) for $\alpha = 98$ for state 0. Time derivative of the cross-correlation $\dot{C}_{0 -1}$ and response functions $\chi_{0 -1}$ demonstrating violation of FDT in (a) and validity of GFDT in (b).
		\label{fig:no_history_alpha_98_0}}
\end{figure}

When examining the second three-state system (Fig.~\ref{fig:3_state_system_model_b}) with broken detailed balance but no adaptation, we found that the GFDT holds as expected. In the main text, we demonstrated the 
necessary correspondence for only one correlation function -- see Fig.~\ref{fig:no_history_alpha_98}. For completeness, here we show the analogous 
results for states $-1$ and $0$ in Figs.~\ref{fig:no_history_alpha_98_neg1} and \ref{fig:no_history_alpha_98_0} respectively. In all of these examples, the 
standard FDT breaks down, but the GFDT relations are valid. 

\section{Linear regime of the equilibrium three-state model}
\label{app-linear}
\begin{figure}
	\includegraphics[width=1\linewidth]{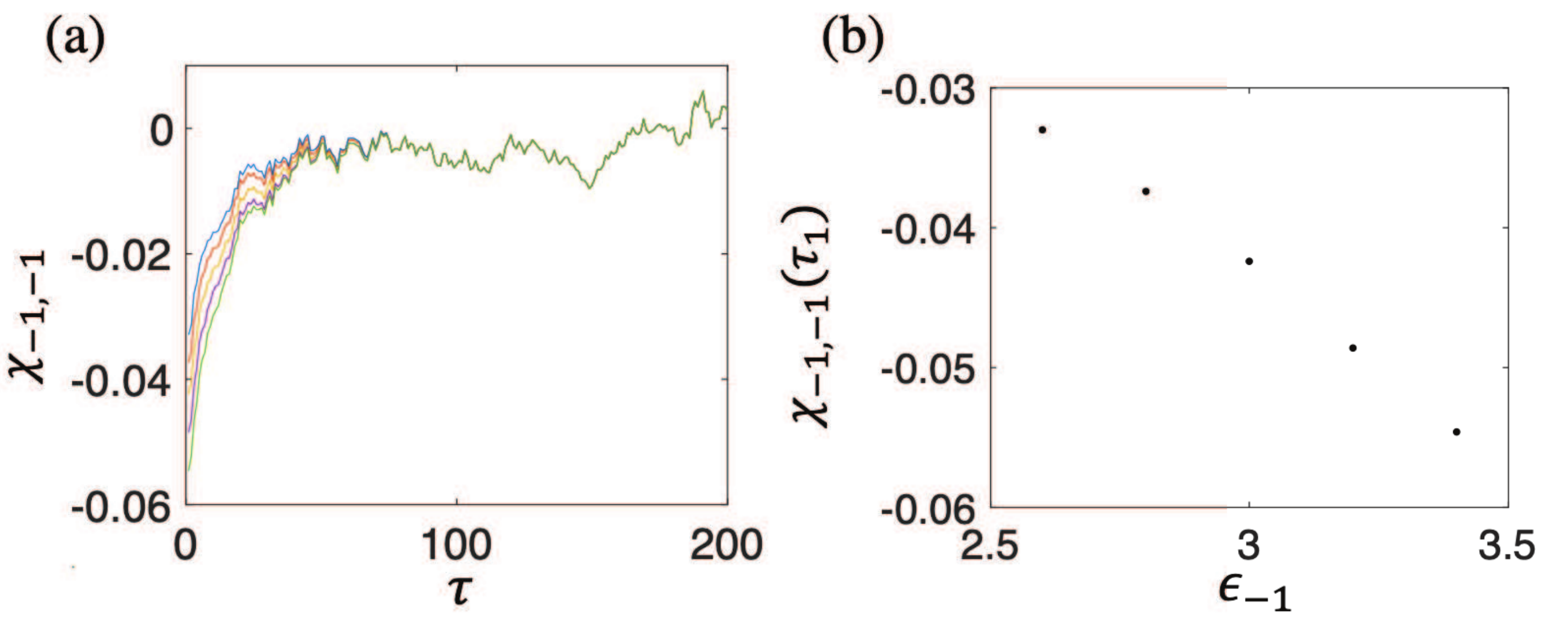}
	\colorcaption{Variation of $\chi_{-1,-1}$ with $\epsilon_{-1}$. (a) With the three-state model obeying detailed balance, we obtain its $\chi_{-1,-1}$ response by setting the energy of state -1, $\epsilon_{-1} = [2.6, 2.8, 3, 3.2, 3.4]$ for one time step. These are respectively colored with blue, orange, yellow, purple and green. (b) The magnitude of the $\chi_{-1,-1}(\tau_1)$ values linearly increase with $\epsilon_{-1}$.
		\label{fig:linear_regime}}
\end{figure}

In Fig.~\ref{fig:no_history_alpha_1}, we perturb the system using an $\epsilon_{-1}$ value of 3. To verify that the response of this forced oscillator is within its linear regime, in panel (a) of Fig.~\ref{fig:linear_regime} we plot over a range of $\epsilon_{-1}$ values (2.6, 2.8, 3, 3.2, and 3.4) their respective $\chi_{-1,-1}$s. Additionally, in panel (b), we show that the magnitude of these response functions at time $\tau_1$ varies with $\epsilon_{-1}$ in a linear fashion.

\section{Adaptively driven three-state system}
\label{app-adaptation-three-state}

\begin{figure}
	\includegraphics[width=1\linewidth]{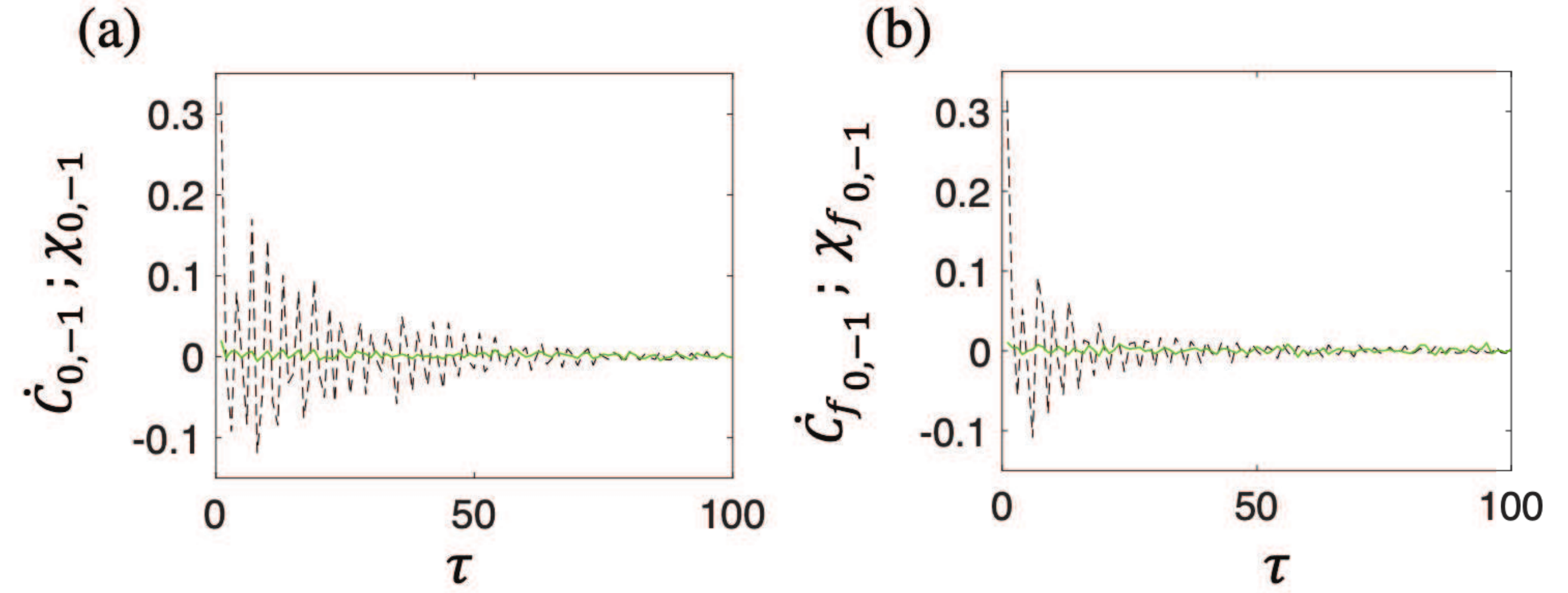}
	\colorcaption{$\dot{C}_{0,-1}$ (black dashed) vs $\chi (\tau)_{0,-1}$ (green solid) for history-dependent $v_{\rm{drift}}$. Time derivative of the cross-correlation function $\dot{C}_{0,-1}$ and the response function $\chi_{0,-1}$ juxtaposed to reveal the breakdown of FDT in (a) and GFDT in (b).
		\label{fig:history_drift_diffusion_state_0}}
\end{figure}

\begin{figure}
	\includegraphics[width=1\linewidth]{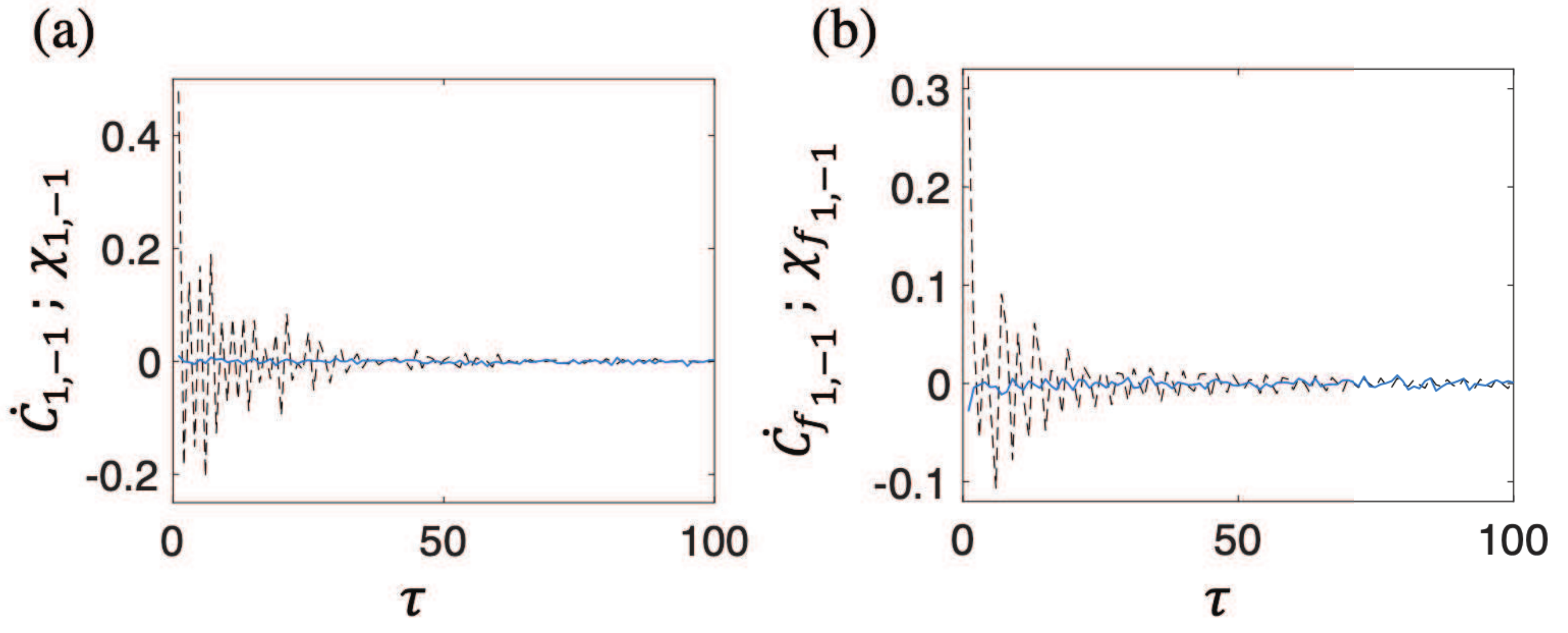}
	\colorcaption{$\dot{C}_{1,-1}$ (black dashed) vs $\chi (\tau)_{1,-1}$ (blue solid) for $v_{\rm{drift}} = 1$. Comparison of the time derivative of the cross-correlation $\dot{C}_{1,-1}$ and response function $\chi_{1 -1}$ exemplifies violation of both the equilibrium (in panel (a)) and generalized (panel (b)) fluctuation-dissipation relations.
		\label{fig:history_drift_diffusion_state_1}}
\end{figure}

In Fig.~\ref{fig:history_drift_diffusion_state_-1}, we depicted the effects of an adaptive drive only for the state -1, namely the violation of both FDT and GFDT.  We obtain similar plots for both states 0 and 1 -- see Figs.~\ref{fig:history_drift_diffusion_state_0} and \ref{fig:history_drift_diffusion_state_1}.

\end{document}